\RequirePackage[colorlinks,allcolors=blue]{hyperref} 
\documentclass[]{agujournal2019}
\usepackage{url} 
\usepackage[inline]{trackchanges} 
\usepackage{soul}

\draftfalse

\journalname{JGR: Planets}

\begin{document}

%
%

\title{Jupiter's Temperate Belt/Zone Contrasts Revealed at Depth by Juno Microwave Observations}

%
%




\authors{L.N. Fletcher\affil{1}, F.A. Oyafuso\affil{2}, M. Allison\affil{3}, A. Ingersoll\affil{4}, L. Li\affil{5}, Y. Kaspi\affil{6}, E. Galanti\affil{6}, M.H. Wong\affil{7}, G.S. Orton\affil{2}, K. Duer\affil{6}, Z. Zhang\affil{4}, C. Li\affil{8}, T. Guillot\affil{9}, S.M. Levin\affil{2}, S. Bolton\affil{10}}


\affiliation{1}{School of Physics and Astronomy, University of Leicester, University Road, Leicester, LE1 7RH, UK}
\affiliation{2}{Jet Propulsion Laboratory, California Institute of Technology, 4800 Oak Grove Drive, Pasadena, CA 91109, USA}

\affiliation{3}{Goddard Institute for Space Studies, New York, NY, USA}

\affiliation{4}{California Institute of Technology, Pasadena, CA, USA}

\affiliation{5}{University of Houston, Houston, TX, USA}

\affiliation{6}{Department of Earth and Planetary Sciences, Weizmann Institute of Science, Rehovot 76100, Israel}

\affiliation{7}{SETI Institute, Mountain View, CA, 94043-5139, USA}

\affiliation{8}{University of Michigan, Ann Arbor, MI, USA.}

\affiliation{9}{Universiti\'{e} C\^{o}te d'Azur, OCA, Lagrange CNRS, 06304 Nice, France}

\affiliation{10}{Southwest Research Institute, San Antonio, Texas, TX, USA.}






\correspondingauthor{Leigh N. Fletcher}{leigh.fletcher@le.ac.uk}




\begin{keypoints}
\item Banded structure of Jupiter's microwave brightness is correlated with the cloud-top winds as far down as 100 bars.
\item Belt/zone contrasts flip sign in the 5-10 bar region, a transition layer coinciding with the water condensation level.
\item Transition can be explained by stacked meridional circulation cells and/or latitudinal gradients in precipitation.
\end{keypoints}


%
%

%
%


\begin{abstract}

Juno Microwave Radiometer (MWR) observations of Jupiter's mid-latitudes reveal a strong correlation between brightness temperature contrasts and zonal winds, confirming that the banded structure extends throughout the troposphere. However, the microwave brightness gradient is observed to change sign with depth: the belts are microwave-bright in the $p<5$ bar range and microwave-dark in the $p>10$ bar range. The transition level (which we call the jovicline) is evident in the MWR 11.5 cm channel, which samples the 5-14 bar range when using the limb-darkening at all emission angles. The transition is located between 4 and 10 bars, and implies that belts change with depth from being NH$_3$-depleted to NH$_3$-enriched, or from physically-warm to physically-cool, or more likely a combination of both. The change in character occurs near the statically stable layer associated with water condensation. The implications of the transition are discussed in terms of ammonia redistribution via meridional circulation cells with opposing flows above and below the water condensation layer, and in terms of the `mushball' precipitation model, which predicts steeper vertical ammonia gradients in the belts versus the zones. We show via the moist thermal wind equation that both the temperature and ammonia interpretations can lead to vertical shear on the zonal winds, but the shear is $\sim50\times$ weaker if only NH$_3$ gradients are considered. Conversely, if MWR observations are associated with kinetic temperature gradients then it would produce zonal winds that increase in strength down to the jovicline, consistent with Galileo probe measurements; then decay slowly at higher pressures.

\end{abstract}

\section*{Plain Language Summary}
One of the core scientific questions for NASA's Juno mission was to explore how Jupiter's famous banded structure might change below the top-most clouds.  Did the alternating bands of temperatures, winds, composition, and clouds simply represent the top of a much deeper circulation pattern?  Juno's microwave radiometer is capable of peering through the clouds to reveal structures extending to great depths, and has revealed a surprise:  belts and zones do persist to pressures of 100 bars or more, but they flip their character at a level which we call the `jovicline,' coinciding with the depths at which water clouds are expected to form and generate a stable layer.  This transition from microwave-bright belts (ammonia depleted and/or physically warm) in the upper layers, to microwave-dark belts (ammonia enriched and/or physically cool) in the deeper layers, and vice versa for the zones, may have implications for the shear on the Jupiter's zonal winds, indicating winds that strengthen with depth down to the jovicline, before decaying slowly at higher pressures.  The origins of the transition is explored in terms of meridional circulations that change with depth, and in terms of models where strong precipitation dominates in the belts.

%
%



\section{Introduction}
\label{intro}

The colourful bands of Jupiter have been the planet's defining characteristic for centuries, discovered mere decades after the invention of the telescope \cite{98hockey}.  The tropospheric bands are organised by east-west zonal jets \cite<e.g.,>{03porco, 06read_jup}, which separate regions exhibiting different temperatures \cite{81pirraglia}, different gaseous composition \cite<e.g., ammonia and phosphine,>{86gierasch, 09fletcher_ph3}, and different aerosol properties \cite<the reflectivity and colour of the clouds and hazes, e.g.,>{04west}.  These bands were historically characterised as high-albedo zones and low-albedo belts, but we adopt a belt-zone nomenclature based on their vorticity.  The zones are anticyclonic and the belts are cyclonic. Zones are cool in the upper troposphere (i.e., adiabatic expansion above the clouds and below the stably stratified tropopause) and have eastward (prograde) jets on their poleward edges, generating potential vorticity gradients that act as barriers to meridional mixing \cite{06read_jup}.  Conversely, belts are warm (adiabatic compression) and feature westward (retrograde) jets on their poleward boundaries.  

The upper-tropospheric belt/zone temperature contrasts encourage condensation of volatiles (e.g., ammonia) in cooler regions, typically producing reflective aerosols in zones and cloud-free conditions in belts, although the correspondence between the zonal jets and the opacity of the clouds \cite<sensed at 5 $\mu$m,>{19antunano} only really holds at low latitudes.  Conversely, the correspondence between the observed cloud-tracked winds and upper tropospheric temperatures persists up to high latitudes near $\pm60^\circ$ \cite{83conrath, 86flasar_jup, 06simon, 16fletcher_texes} and implies, via the thermal wind equation \cite{04holton}, that the zonal jets decay with altitude from the cloud-tops to the tropopause \cite{81pirraglia, 90conrath}.  The source of the dissipative mechanism causing this decay with height remains unclear and has never been directly observed, but could be related to wave or eddy stresses opposing the winds \cite{89pirraglia, 93orsolini}.  Finally, the latitudinal distribution of chemicals such as ammonia \cite{86gierasch, 06achterberg, 16depater, 17li}, phosphine \cite{09fletcher_ph3, 17giles, 20grassi}, and para-hydrogen \cite{98conrath, 17fletcher_sofia}, combined with the observed temperature and aerosol distributions, suggest that the atmospheric circulation in the upper troposphere is dominated by rising motions over zones, zone-to-belt meridional transport at high altitude, and sinking over the belts.  This is the ``classical'' picture of belt/zone circulation envisaged by \citeA{51hess} and \citeA{76stone}, and is often likened to `Hadley-like' circulations in the terrestrial atmosphere, whereby warm tropical air rises and moves poleward (a thermally-direct circulation), being deflected eastward by the Coriolis effect to generate sub-tropical jet streams.

Insights from Voyager, Galileo, and Cassini have challenged this conceptual picture, as reviewed by \citeA{20fletcher_beltzone}.  Lightning was detected as optical flashes \cite{99little, 00gierasch, 07baines}, and was found to be prevalent in the belts but either absent or obscured in the zones.  This suggested moist air converging and rising in the belts, potentially in narrow convective plumes embedded within regions of net subsidence \cite{87lunine,00ingersoll, 05showman}.  Furthermore, cloud-tracking by Voyager \cite{81ingersoll} and Cassini \cite{06salyk} identified eddies converging and supplying momentum to the eastward jets, via a process analogous to Earth's Ferrel cells \cite{06vallis}.  This forcing of the jets by flux convergence can be confined to shallow layers within the clouds and yet still produce jets that extend deep \cite{08lian}.  However, the forcing must be balanced by a compensating meridional flow, which has rising motions in belts, belt-to-zone meridional transport, and sinking over the zones.  Such a belt/zone circulation is opposite to that postulated for the upper troposphere, and has led to a hypothesis of `stacked circulation cells,' with deep Ferrel-like cells dominated by eddy-forcing of the zonal winds, and upper cells of eddy-dissipation and wind decay \cite{00ingersoll, 05showman, 20fletcher_beltzone}, with a poorly defined transition somewhere within the `weather layer' above the water clouds.  Such counter-rotating stacked cells have been observed in numerical simulations with prescribed heating and eddy momentum fluxes \cite{05yamazaki, 09zuchowski}, and general circulation models (GCMs) hint at changes to the magnitude of eddy-momentum flux convergence as a function of altitude \cite{18young, 20spiga}.  

Juno's exploration of Jupiter provides an opportunity to explore belt/zone contrasts below the cloud tops, and to test the stacked-cell hypothesis.  Jupiter's winds have been found to extend to approximately 3000 km below the clouds \cite{18kaspi,18guillot}, to the level where Ohmic dissipation may become important \cite{08liu, 17cao, 20kaspi,20galanti}.  The slow decay with depth suggests that the meridional temperature gradients must be weak but opposite to that seen in the upper troposphere (where winds strengthen with depth).  Observations by Juno's microwave radiometer (MWR) found the vertical distribution of ammonia to be variable across latitudes from $40^\circ$S to $40^\circ$N, with widespread depletion down to 40-60 bar \cite<perijove 1, 27 August 2016,>{17bolton, 17li, 17ingersoll}.  Previously, the ammonia cross-section was observed to be dominated by an NH$_3$-rich column at the equator, flanked by NH$_3$-depleted belts evident in both the mid-IR \cite{06achterberg, 16fletcher_texes} and ground-based millimetre and sub-millimetre observations \cite{16depater}.  Although some form of NH$_3$ depletion might result from precipitation \cite{17ingersoll}, it was a challenge to get this below the 10-bar level \cite{19li} without invoking a process using robust `mushballs' \cite{20guillot_mushball} composed of mixed-phase ammonia/water condensates \cite{73weidenschilling}.  From these Juno microwave observations in 2016, \citeA{17ingersoll} noted that the correlation of ammonia variations with the belts and zones was rather weak at $p<2$ bars, but that the correlation was better from $p=40$ to $60$ bars, where the belts have higher ammonia abundances than the zones, opposite to what was seen in the upper troposphere.  The very existence of localised NH$_3$ anomalies suggests that upwelling and subsidence must be occurring in the presence of a vertical NH$_3$ gradient throughout the range of MWR sensitivity.  Furthermore, \citeA{20duer} used these same PJ1 data to reveal correlations between cloud-top winds and the NH$_3$ abundances and concentration gradients, supporting the inference of meridional circulation cells in the altitude range sounded by MWR.  Finally, observations from the Very Large Array in 2014 \cite<VLA, probing as deep as $\sim7$ bar at 10 cm, >{19depater_vla} also tentatively suggested a brightness temperature reversal for a single band near the $21^\circ$N jet, but this was for a single location and a shallower pressure than the phenomenon identified in our study.

In this study, we investigate the correlation between Jupiter's cloud-top winds and microwave brightness using observations spanning the first two years of Juno operations (2016-2018), focusing on the mid-latitude temperate domains away from the strong NH$_3$ gradients at the equator (Section \ref{datacorr}).  We report the existence of a level at which the microwave brightness contrasts reverse, which we call the `jovicline' via analogy to terrestrial oceanography.  By exploiting the emission-angle dependence of the brightness temperatures to sound a range of altitudes, we show in Section \ref{transition} how we constrain the pressure of the transition between microwave-bright belts in the upper troposphere, and microwave-dark belts in the deeper atmosphere.  We aim to show, in a model-independent way, that the transition is evident from the data alone, irrespective of its interpretation.  Section \ref{discuss} shows how the identification of this transition relates to atmospheric temperatures, winds, and ammonia within the stacked-cell hypothesis, and explores alternative scenarios for the observed contrasts.

\section{Juno Microwave Contrasts}
\label{datacorr}

\begin{figure*}
\begin{center}
\includegraphics[angle=0,width=1.2\textwidth]{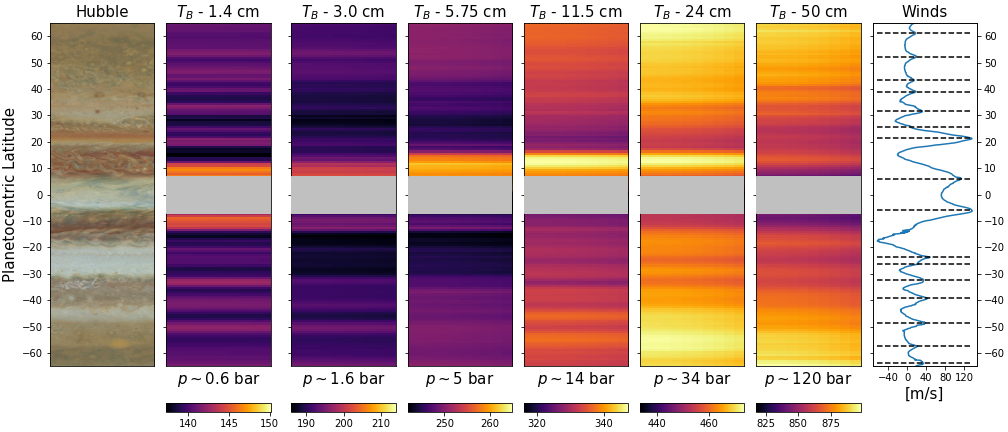}
\caption{Averaged nadir brightness temperatures for each of the six MWR channels as a function of planetocentric latitude \cite{20oyafuso}, compared to Hubble/WFC3 observations acquired on 2017-04-03 \cite<left, for a randomly selected longitude, >{20wong} and cloud-tracked zonal winds \cite<right, with eastward jets indicated as horizontal dashed lines>{03porco}. The microwave-dark equator \cite{17li} has been omitted from the MWR images (grey boxes equatorward of $\pm7^\circ$) to reveal the contrasts at mid-latitudes.  Note the transition from microwave-bright bands in shallow-sounding channels (1.4-5.75 cm) to microwave-dark band in deep-sounding channels (11.5-50 cm).  Estimated pressures for each channel are discussed in Section \ref{contribution}.  }
\label{beltzone}
\end{center}
\end{figure*}

\subsection{MWR Observations}
In this section we demonstrate the correlation between microwave brightness temperature gradients and the locations of Jupiter's cloud-tracked zonal jets, as shown in Fig. \ref{beltzone}.  The Microwave Radiometer \cite<MWR,>{17janssen} is part of a suite of remote sensing instruments on the Juno spacecraft \cite{17bolton}, which has been in a 53-day polar orbit around Jupiter since July 2016.  The elliptical orbits bring the spinning spacecraft within 3000-4000 km of the jovian cloud tops during the $\sim2$-hour pole-to-pole perijove (PJ) passes, during which time the fields-of-view of the six MWR receivers (spanning 0.6-21.9 GHz, or 1.4-50 cm) are swept over the scene.  MWR measurements provide two key capabilities over previous ground-based radio measurements; (1) they are able to unambiguously separate Jupiter's synchrotron emission from atmospheric thermal emission, particularly important for observations at $p>5$ bars, and (2) the 2-rpm spin of the spacecraft allows a direct measurement of brightness as a function of emission angle for each position, which will be key to this study of the belt/zone transition. 

\citeA{20oyafuso} describe how the jovian brightness temperatures, $T_B$, are deconvolved from the antenna temperatures, removing the galactic and synchrotron backgrounds and accounting for the antenna beam pattern and contributions from sidelobes (a feature of the beam pattern).  The result is a $T_B$ as if it were measured along a narrow pencil-beam targeting a particular latitude $\phi$ (sampled on a grid of 255 points from pole to pole) and emission angle.  The dependence of the brightness on the emission-angle cosine $\mu$ is known as the limb darkening, and is expressed via the quadratic function \cite{20oyafuso}:
\begin{equation}
        T_B(\mu)=\xi(\mu)\left[c_0 - c_1\frac{1-\mu}{1-\mu^*} + \frac{c_2}{2}\frac{(\mu-\mu^*)(1-\mu)}{(1-\mu^*)^2}\right]
\label{eq:coeff}
\end{equation}
where $\mu^*$ is set to 0.8; the coefficient $c_0$ is the nadir brightness temperature ($\mu=1.0$), $c_1$ is the absolute limb darkening when $\mu=\mu^*=0.8$ (chosen to correspond to an emission angle of $37^\circ$), and $c_2$ represents a further decline in brightness at $53^\circ$ ($\mu=0.6$) beyond that obtained from a linear extrapolation from nadir to $37^\circ$.  The range of $\mu$ between 1.0 and 0.6 was selected as the most appropriate for the MWR emission angle coverage.  The parameter $\xi(\mu)$ is a shape function that accounts for imperfections in the quadratic fit to the limb-darkening dependence beyond $53^\circ$ \cite<see>[for full details]{20oyafuso}.  

This work uses $T_B(\phi,\mu)$ reconstructed from the fitted coefficients in equation \ref{eq:coeff} for each latitude from PJ1 (27 August 2016) through PJ12 (1 April 2018).  Data from PJ10 (December 2017) and PJ11 (February 2018) were not used because the spacecraft orientation was optimised for gravity science (i.e., favouring continuous Earth pointing), and no data were obtained during PJ2 (October 2016).  MWR samples narrow longitudinal swaths during each of the nine selected perijoves, which are used to represent Jupiter's zonally-averaged microwave brightness.  However, to filter out coefficients that resulted from poor quality quadratic fits to the observed limb darkening, we construct a weighted average of each coefficient at each latitude, weighting by (i) a local $\chi^2$ describing the goodness of fit of the quadratic in equation \ref{eq:coeff} to the $T_B(\phi,\mu)$ measurements; and by (ii) a spatial contribution function that weights by the square root of the effective number of measurements at a given latitude \cite<see>[for details]{20oyafuso}.

\begin{figure*}
\begin{center}
\includegraphics[angle=0,width=0.9\textwidth]{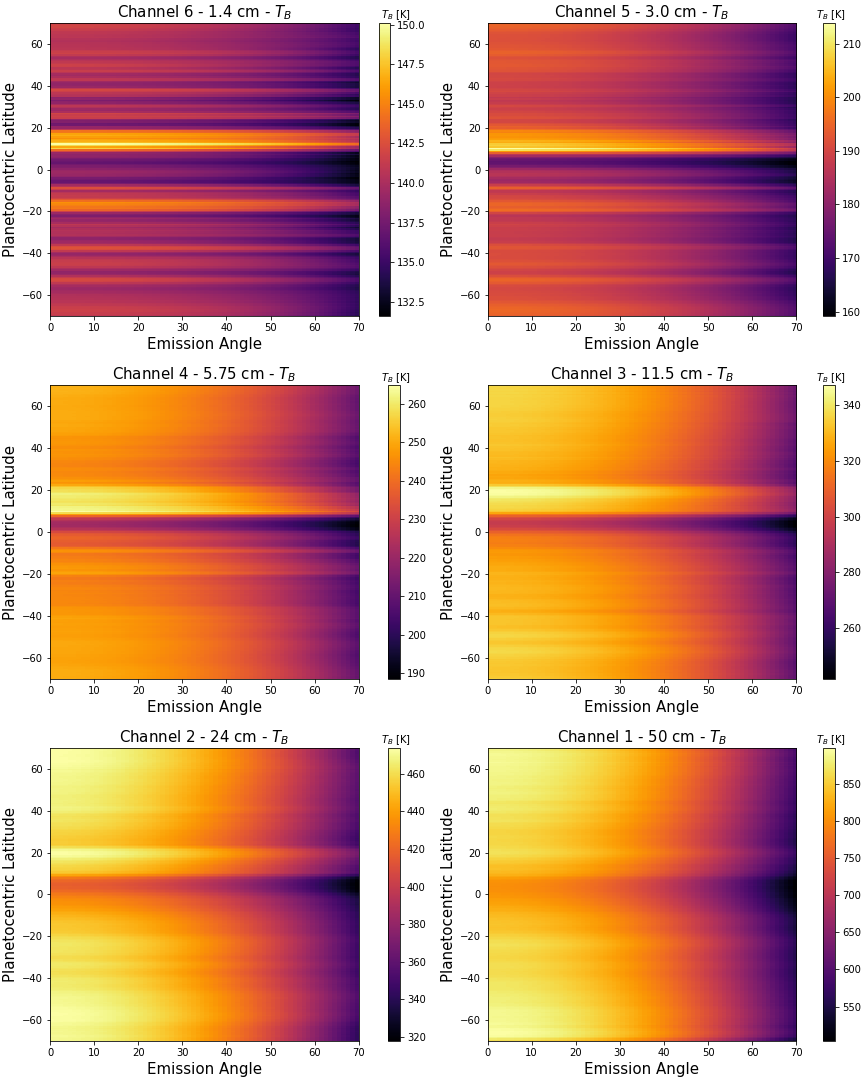}
\caption{Deconvolved brightness temperatures as a function of emission angle and planetocentric latitude, formed from a weighted average of nine Juno perijoves between August 2016 and April 2018.  Banded structure is observed in all channels, but the contrast is dominated by the tropics.  No attempt has been made to remove the latitudinal dependence of $T_B$ on atmospheric scale height (which depends on effective gravitational acceleration), see Section \ref{nadirbrightness}. }
\label{limbdarken}
\end{center}
\end{figure*}

The weighted-average nadir $T_B(\phi)$, using only the $c_0$ coefficient from Eq. \ref{eq:coeff}, is shown in Fig. \ref{beltzone} and compared to a colour map from Hubble/WFC3 \cite<acquired on 2017-04-03, shortly after PJ5,>{20wong}, and compared to Jupiter's cloud-top zonal winds \cite{03porco}.  In addition, the weighted average $T_B(\phi,\mu)$ is shown in Fig. \ref{limbdarken}, showing the limb darkening for each of the six channels.  These figures reveal a banded structure at all pressure levels sampled by these data, from $\sim120$ bars at 50 cm (Channel 1) to $\sim0.6$ bars at 1.4 cm (Channel 6).  The percentage limb darkening at $45^\circ$ emission angle ranges from 1\% at 0.6 bars (i.e., minimal limb darkening) to 13-15\% at 100 bars (strong limb darkening), consistent with \citeA{20oyafuso}.  No attempt is made in Fig. \ref{limbdarken} to adjust for the poleward increase in brightness resulting from the change in Jupiter's atmospheric scale height, which depends on effective gravitational acceleration (see Section \ref{nadirbrightness}).  The tropical contrasts between the microwave-dark Equatorial Zone (EZ, $6^\circ$N-$6^\circ$S) and the microwave-bright North/South Equatorial Belts (NEB $6.0-15.2^\circ$N and SEB $6.0-17.4^\circ$S) dominate Fig. \ref{limbdarken} at all pressure levels, interpreted by \citeA{17li} and \citeA{17ingersoll} as a column of enriched NH$_3$ gas at the equator, with strong NH$_3$ depletion over the neighbouring belts.  For our purposes, these strong tropical contrasts dominate the colour scale in Fig. \ref{limbdarken} and render the mid-latitude belt/zone contrasts harder to see, so we show the nadir $T_B$ polewards of $\pm20^\circ$ latitude (i.e., the $c_0$ coefficients of Eq. \ref{eq:coeff}) in Fig. \ref{nadirTB}, to be discussed in the next section.

\subsection{Nadir Brightness Gradients}
\label{nadirbrightness}

\begin{figure*}
\begin{center}
\includegraphics[angle=0,width=0.9\textwidth]{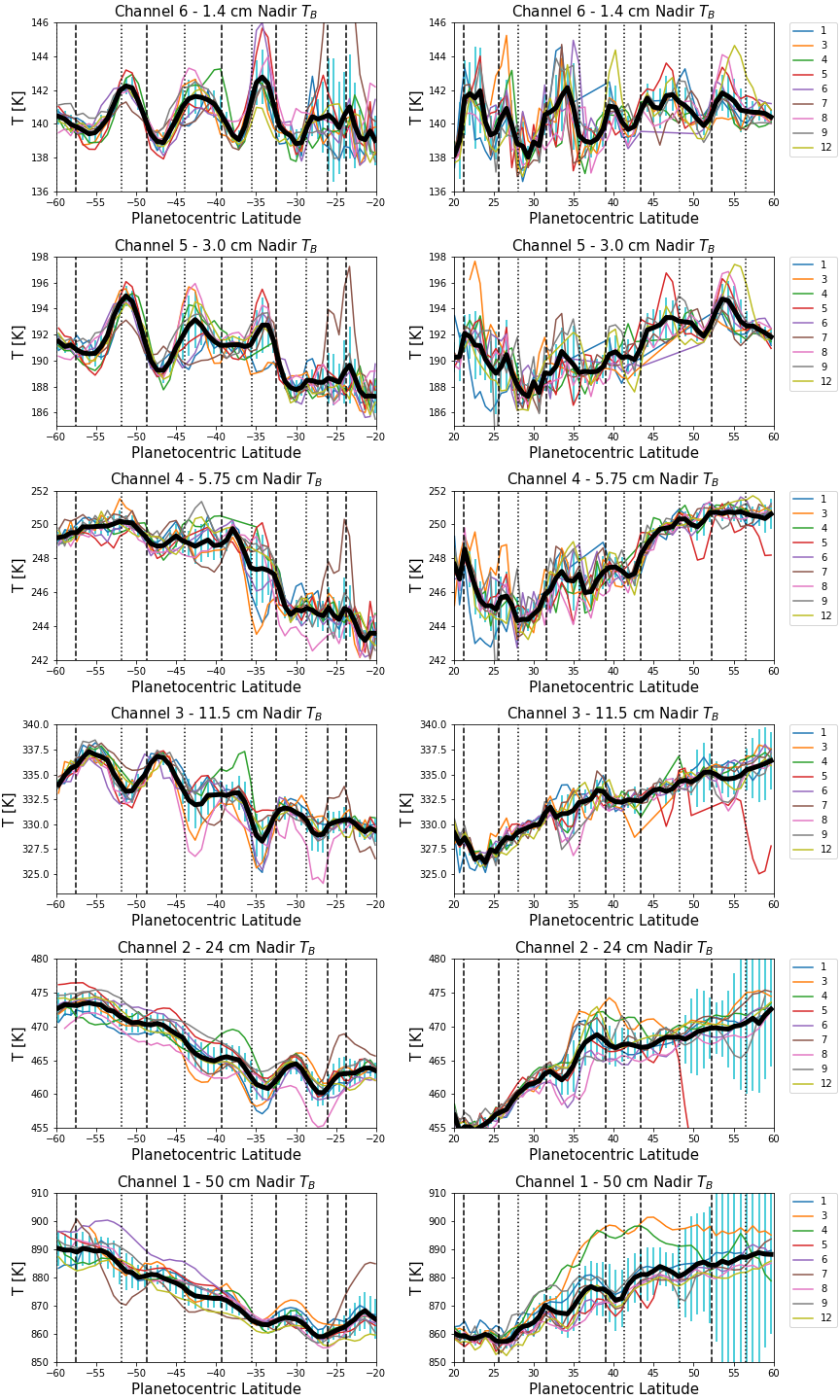}
\caption{Nadir microwave brightness temperatures for all nine perijoves (coloured lines) compared to the weighted average (thick black line) to show the filtering process.  Uncertainties on the weighted average are shown by the blue bars, indicating discrepancies between perijoves. These are compared to the peaks of eastward (dashed) and westward (dotted) zonal winds as measured by Cassini \cite{03porco}.  Note that uncertainties become large at high northern latitudes for wavelengths longer than 11.5 cm, due to the introduction of synchrotron noise into the beam.}
\label{nadirTB}
\end{center}
\end{figure*}

Fig. \ref{nadirTB} demonstrates how the filtering process of \citeA{20oyafuso} identifies measurements that appear to differ substantially from other perijoves.  For example, the microwave-bright southern periphery of the Great Red Spot was observed on PJ7 \cite{17li_grs} and is a significant outlier near $25^\circ$S, but the poor goodness-of-fit ($\chi^2$) for the quadratic in Eq. \ref{eq:coeff} for these latitudes means that PJ7 does not contribute significantly to our average.  Similarly, a screening algorithm is used to remove observations contaminated by synchrotron emission, meaning that there will be fewer measurements available in affected latitudes for the quadratic fitting \cite<see Section 2.5 of>{20oyafuso}.  This was particularly true for PJ3 and PJ4 at northern mid-latitudes, which appear anomalously bright but are constrained by very few uncontaminated measurements, such that their reduced weighting via the spatial contribution function minimises their contribution to the weighted average.  The thick black line shows our best estimate of the microwave banding \cite<consistent with>{20oyafuso}, and is compared to the locations of the eastward (prograde, dashed) and westward (retrograde, dotted) jets as determined by Cassini/ISS cloud-tracking of zonal winds $u$ \cite{03porco}, extracted via identifying locations where the vorticity $-\partial{u}/\partial{y}=0$ (where $y$ is the north-south distance in kilometres, accounting for the radius of curvature for an oblate spheroid).  Similar calculations using Hubble cloud-tracked winds in 2017-19 are shown in \ref{appS1}, but the location of the jets has not changed significantly with time \cite{17tollefson, 20wong}.  We use these velocity minima and maxima in Fig. \ref{beltzone} to define the locations of Jupiter's cloud-top belts and zones, rather than the aerosol opacity, colour, and reflectivity, which are not good proxies for the underlying zonal wind structure \cite{20fletcher_beltzone}.  



\begin{figure*}
\begin{center}
\includegraphics[angle=0,width=0.9\textwidth]{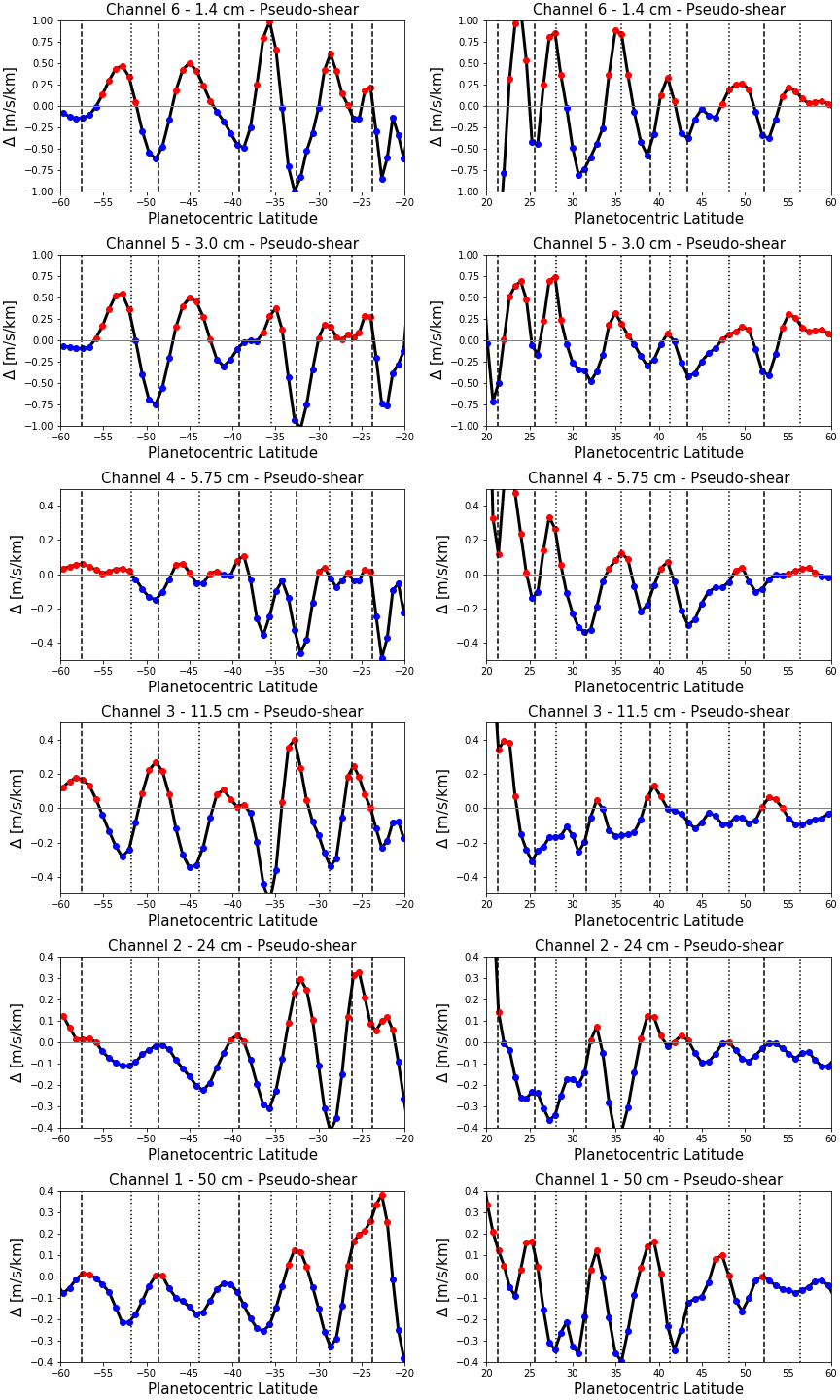}
\caption{Nadir microwave brightness gradients for temperate latitudes, corrected by both the Coriolis parameter and gravitational acceleration to represent `pseudo-shear' in m/s/km.  Regions of negative pseudo-shear are represented by blue points, regions of positive pseudo-shear are represented by red points.  These are compared to the peaks of eastward (dashed) and westward (dotted) zonal winds as measured by Cassini \cite{03porco}.}
\label{nadir_dudz}
\end{center}
\end{figure*}

To better emphasise the gradients observed by MWR, we convert the $T_B$ measurements into a `pseudo-shear' $\Delta$ by analogy to the thermal wind equation \cite{04holton}, assuming \textit{constant pressure surfaces}:
\begin{equation}
    \Delta=-\frac{g}{fT_B}\frac{\partial T_B}{\partial y} 
\end{equation}
where we replace the kinetic temperature of the atmosphere with the brightness temperature. $f$ is the Coriolis parameter, $g$ is the gravitational acceleration at the particular pressure and latitude, and the brightness temperature derivative is evaluated on isobars (constant-pressure surfaces). At this stage, we make no connection between $\Delta$ and the shear on the zonal jets, but use this formalism simply to denote the edges of the microwave belts and zones.  We plot $\Delta$ in Fig. \ref{nadir_dudz}, showing how the peaks in the microwave brightness gradients are co-located with the cloud-tracked zonal jets (the strength of the correlation will be explored below).  Dashed lines are eastward jets (zones on the equatorward sides, belts on the poleward sides); dotted lines are westward jets (zones on the poleward side, belts on the equatorward side).  Blue points are used to denote a negative gradient, red points are used for a positive gradient, and the patterns provide our first sign that a transition in belt/zone gradients occurs between the deep-sensing channels 1-3 (6 to greater than 100 bars), and the shallow-sensing channels 4-6 (0.6 to 5.0 bars).

We can see this reversal in $\Delta$ by tracking single jets in Fig. \ref{nadir_dudz}.  For example, the prograde jets at $48.6^\circ$S and $32.5^\circ$S coincide with local minima of negative $\Delta$ in the 0.6-5.0 bar range, but flip to being local maxima of positive $\Delta$ in the 10-100 bar range.  Conversely, the retrograde jets at $35.5^\circ$S and $43.9^\circ$S coincide with local maxima of positive $\Delta$ at shallow depths, and local minima of negative $\Delta$ at deeper levels.  This reversal in $\Delta$ has the effect of transitioning a traditional jovian belt (with prograde jets on their equatorward edges) from microwave-bright at shallow levels to microwave-dark at deeper levels, and vice versa for zones (with prograde jets on their poleward edges), as previously identified in PJ1 observations between $40^\circ$S and $40^\circ$N by \citeA{17ingersoll}.  The correspondence between $\Delta$ and the cloud-tracked winds is not perfect, and we explore the statistical significance of the correlations in Section \ref{correlation}.  In particular, we caution that (i) the correspondence is clear in the south but only suggestive (at best) in the north; and (ii) a residual equator-to-pole gradient remains in the data as a shift towards negative values of $\Delta$ in the deep-sounding channels 1-3.  The origin of this deep poleward gradient of deep temperature and/or NH$_3$, superimposed onto the banded structure, is the topic of an ongoing investigation.

We omitted latitudes smaller than $\pm20^\circ$ from Figs. \ref{nadirTB}-\ref{nadir_dudz}.  However, the $\Delta$ reversal is prominent for the retrograde NEBn and SEBs jets at $15.2^\circ$N and $17.4^\circ$S, respectively (from positive $\Delta$ at shallow depths, to negative $\Delta$ at deeper levels).  This can be seen in Fig. \ref{limbdarken}, where an extremely bright band is observed in deep-sensing Channels 1-3 in the $15.2-21.3^\circ$N region (the North Tropical Zone, NTrZ), but not in shallow-sensing Channels 4-6.  Right at the equator, the prograde jets bounding the EZ (the NEBs at $6.0^\circ$N and the SEBn at $6.0^\circ$S) are the only jets where no $\Delta$ reversal is observed, it remains negative at all levels given that the equatorial zone is always microwave-dark in Fig. \ref{limbdarken}.   This is consistent with the EZ being an unusual region of elevated NH$_3$ abundance \cite{17li}, and what follows focuses on the banded structure away from the equatorial belts and zones.   

Finally, the Cassini/ISS winds (shown later in Fig. \ref{winds_xsection}) show the existence of small notches in the $\partial{u}/\partial{y}$ profiles near $26.1^\circ$S and $25.6^\circ$N.  We have treated these as additional eastward jets in Fig. \ref{nadir_dudz}, although this is not standard nomenclature (they exist in the middle of the NTB and STB, respectively).  The STB wind feature appears to be strong adjacent to the `structured sectors' known as the STB Ghost, Spectre, and other dark segments \cite{20inurrigarro}, and absent elsewhere (J. Rogers, \textit{pers.comms.}). The NTB feature could be sub-dividing the belt in two. However, MWR reveals that there are substantial brightness gradients ($\Delta$, with a reversal in sign) associated with both of these features in each channel, suggesting that they are more important to the flow field than suggested by the cloud-tracked winds.  These additional `mid-temperate-belt' jets will be the subject of future investigations.

\subsection{Correlation Analysis}
\label{correlation}

\begin{figure*}
\begin{center}
\includegraphics[angle=0,width=1.1\textwidth]{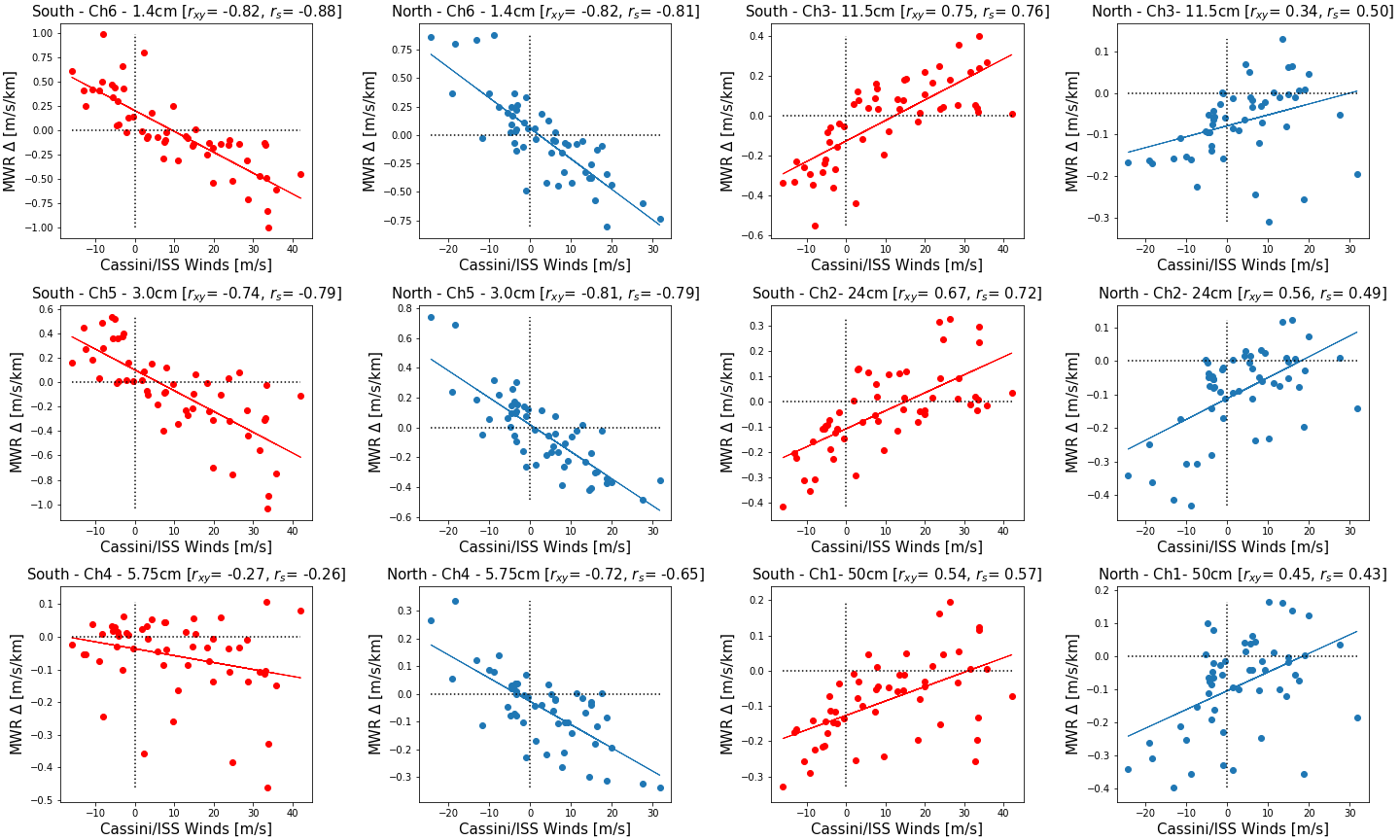}
\caption{Scatter plots revealing negative (channels 4-6, left columns) and positive (channels 1-3, right columns) correlations between the nadir microwave $T_B$ gradients $\Delta$ and the Cassini cloud-tracked winds.  Only latitudes between $25^\circ$ and $65^\circ$ in each hemisphere are included.  Southern-hemisphere correlations are in red, northern-hemisphere correlations are in blue.  A linear trend line has been added as a guide. The Pearson $r_{xy}$ and Spearman's ranked $r_s$ correlation coefficients are provided for each channel and hemisphere.  See \ref{appS1} for similar scatter plots computed using Hubble winds in 2017-19 \cite{20wong, 17tollefson}. }
\label{scatter}
\end{center}
\end{figure*}

In Section \ref{nadirbrightness} we noted that the correlations between the cloud-top winds and the microwave brightness gradients, $\Delta$, were not perfect.  Fig. \ref{scatter} provides a scatter plot of the nadir $\Delta$ versus the Cassini/ISS cloud-top winds for the northern ($25-65^\circ$N) and southern ($25-65^\circ$S) hemispheres, for all six channels.  We restrict this analysis to temperate mid-latitudes $>\pm25^\circ$, excluding Jupiter's fastest retrograde jet (the SEBs at $17.4^\circ$S) and the fastest prograde jet (the NTBs at $21.3^\circ$N) as their extreme speeds would otherwise dominate the correlation analysis, and discuss the importance of these asymmetric jets later in Section \ref{comparegrav}.  As expected from the comparison of $\Delta$ with the jet peaks in Fig. \ref{nadir_dudz}, the scatter plots fall into two groups:  deep-sounding channels (1-3, 11.5-50 cm sounding 10-100 bars) with a positive correlation between prograde velocities and $\Delta$, and shallow-sounding channels (4-6, 1.4-5.75 cm, sounding 0.6-5.0 bars) with negative correlation between prograde velocities and $\Delta$.

Fig. \ref{scatter} shows qualitatively that (i) channel 4 (5.75 cm) shows the weakest correlation in the south, but channel 3 (11.5 cm) shows the weakest correlation in the north; and (ii) the correlations look generally stronger in the south than the north.  To quantify this, we compute the Pearson correlation coefficient ($r_{xy}$, measuring the linear correlation between the winds and $\Delta$) and the Spearman rank correlation coefficient ($r_s$, assessing the strength of the link between the two parameters), and record them in Fig. \ref{scatter}.  We also compute the probability values (p-values) for each correlation, with values significantly smaller than 0.05 allowing us to firmly reject the null hypothesis that the winds and $\Delta$ are uncorrelated (these are provided in Tables \ref{S1} and \ref{S2}).  Confirming the qualitative assessment in Fig. \ref{scatter}, p-values are smallest (and the correlation is highly statistically significant) for channel 5-6, and highest but still significant ($\sim0.01$) for channel 4.  We also computed these correlations using Hubble-derived zonal wind fields in 2017 \cite{17tollefson} and 2019 \cite{20wong}, finding small improvements to the correlation without changing the conclusions - these computations can be found in \ref{appS1}.

The strength of the correlation depends on which perijoves are included in our weighted average, and which latitudes we include in the figure.  In \ref{appS2} we test the robustness of the correlations by selecting random pairs of perijoves from the nine studied here, recomputing the correlation coefficients and $p$-values for each pair and showing that the correlation remains significant, as it was when it was first noted in PJ1 data (August 2016) \cite{17ingersoll, 20oyafuso}.  We also recomputed the correlation coefficients assuming winds that varied along cylinders parallel to the rotation axis \cite{20duer}, and found negligible changes to the strength of the correlations observed in Fig. \ref{scatter}.

\begin{figure*}
\begin{center}
\includegraphics[angle=0,width=0.8\textwidth]{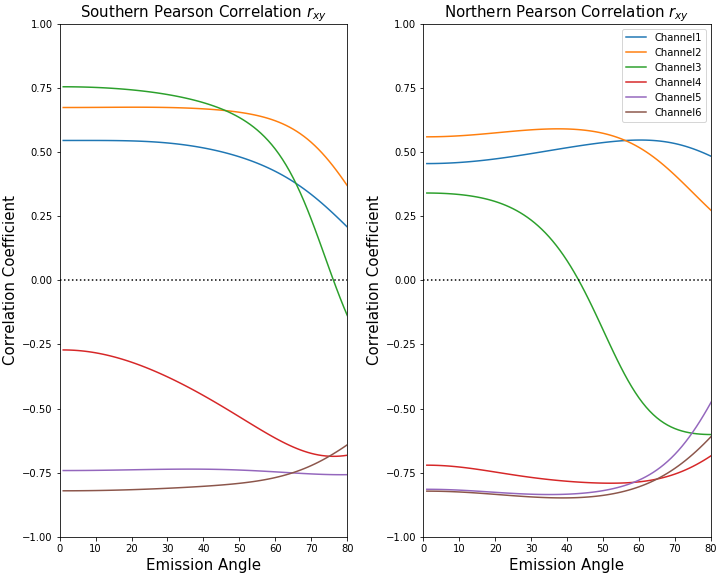}
\caption{Linear correlation between microwave $T_B$ gradients ($\Delta_\mu$) and cloud-top winds calculated on a $1^\circ$ grid at all emission angles (see Section \ref{correlation} for a discussion of reliability at emission angles exceeding $\sim60^\circ$).  The channels naturally fall into two groups (positive and negative correlations), with a cross-over in Channel 3.  These coefficients are hemispheric averages over the $25-65^\circ$ latitude ranges. }
\label{pearson}
\end{center}
\end{figure*}

Finally, we can extend the nadir-only analysis of Fig. \ref{scatter} to all emission angles sampled by MWR, and represented by the limb-darkened brightness temperatures in Fig. \ref{limbdarken}.  We now calculate $\Delta_\mu$ for all $T_B(\phi,\mu)$ values (the $\mu$ subscript denotes that we now include all emission angles), and recompute the Pearson $r_{xy}$ in Fig. \ref{pearson}.  The six channels still naturally fall into two groups - negative correlation at shallow depths, positive correlation at deeper levels.  But Fig. \ref{pearson} also shows that the transition from positive to negative correlation \textit{occurs within a single channel}, channel 3 (11.5 cm), near $45^\circ$ emission angle in the north, and $75^\circ$ emission angle in the south, although we stress that these are averages over all the jets in the $25-65^\circ$ latitude ranges in both hemispheres.  As contribution functions shift higher with increasing emission angle, this provides a rough estimate of the transition pressure as being somewhere between the 14-bar level sounded in channel 3 and the 5-bar level sounded by channel 4.  However, we caution that the deconvolution process of \citeA{20oyafuso} avoided contributions from emission angles exceeding $53^\circ$, such that the southern hemisphere $75^\circ$ crossover in channel 3 depends somewhat on our choice of functional form to represent the limb darkening (Eq. \ref{eq:coeff}).  This should be considered at the edge of the MWR capabilities (i.e., the cross-over happens somewhere between the depths sensed by channels 3 and 4), whereas the northern hemisphere crossover in channel 3 is more convincing.  Indeed, the channel-3 switch from weak positive correlation at nadir ($r_{xy}=0.34$, $p_{xy}=1\times10^{-2}$) to slightly stronger negative correlation at $60^\circ$ emission angle ($r_{xy}=-0.46$, $p_{xy}=5\times10^{-4}$) in Fig. \ref{pearson} is statistically significant.  Fig. \ref{pearsonp} shows how these $p_{xy}$ values vary with emission angle. In Section \ref{transition}, we use the limb-darkening dependence to refine the altitude of the transition point.

\section{Assessing the Transition Depth}
\label{transition}

The MWR data presented in the previous section demonstrated the existence of a transition in the sign of the microwave $T_B$ brightness gradients ($\Delta$), somewhere within the 5-14-bar region sounded by Channels 4 and 3.  This could be seen directly from the deconvolved MWR observations, using the limb-darkening coefficients extracted using the techniques in \citeA{20oyafuso}.  The identification of this transition is independent of any radiative transfer modelling for emission angles smaller than $53^\circ$.  However, the shape function in Eq. \ref{eq:coeff} \cite<estimated from the discrepancy between modelled limb-darkening and the simple polynomial fits,>[]{20oyafuso} begins to deviate from unity beyond $53^\circ$, introducing some weak model dependence to the deconvolved MWR observations at the highest angles.  Further constraints on the altitude of the transition requires an estimation of the angular dependence of MWR contribution functions at each wavelength.  We will use the contribution functions to assign each measured $T_B$ to an estimated pressure level.

\subsection{MWR Contribution Functions}
\label{contribution}
\begin{figure*}
\begin{center}
\includegraphics[angle=0,width=1.0\textwidth]{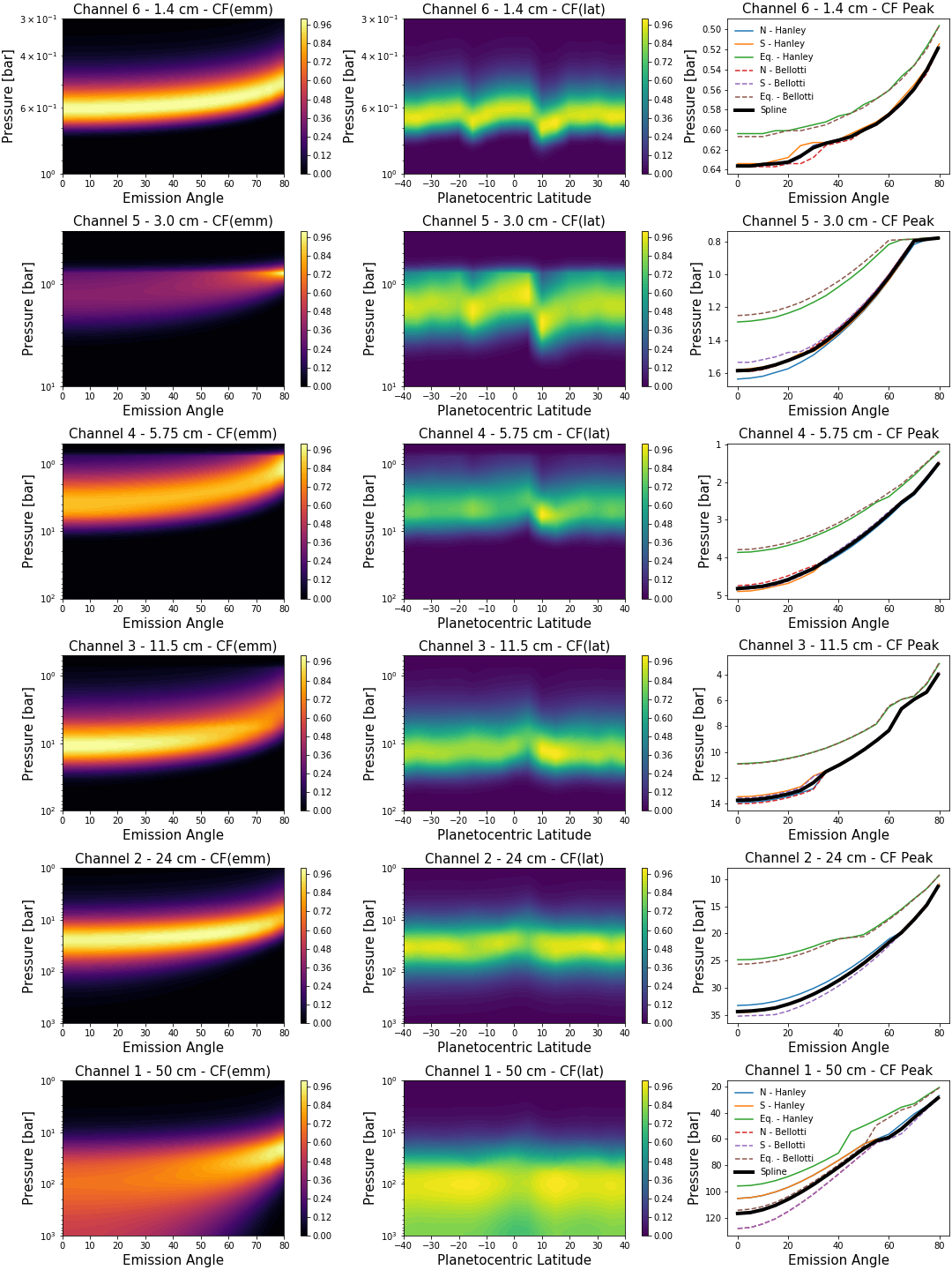}
\caption{Contribution functions based on the retrieved distribution of NH$_3$ versus latitude and pressure based on \citeA{20guillot_ammonia}, with a modified NH$_3$ gradient at $p<0.6$ bars to remove a discontinuity. Left:  normalised contribution functions as a function of emission angle for the equator.  Centre: normalised contribution functions at zero emission angle (nadir view) for all latitudes.  Right: peak pressure of the contribution function averaged over three regions (north $20^\circ$N to $40^\circ$N; south $20^\circ$S to $40^\circ$S; and equator $5^\circ$N to $5^\circ$S) using two different NH$_3$ opacity models - \citeA{09hanley} as the solid lines and \citeA{16bellotti} as the dashed lines. The solid black line is the spline-interpolated contribution function described in the main text.  }
\label{contfn}
\end{center}
\end{figure*}

We use the \textit{Jupiter Atmospheric Radiative Transfer Model} \cite<JAMRT,>[]{17janssen} to calculate the dependence of the contribution function on emission angle, as shown in Fig. \ref{contfn}.  Instead of using the standard JAMRT model with a lower boundary condition of 351 ppm of NH$_3$ \cite<equivalent to $2.76\times$ protosolar ammonia,>[]{20li_water}, and an NH$_3$ profile declining with height due to equilibrium cloud condensation (see \ref{appS3}), we instead use the retrieved NH$_3$ distribution on a $5^\circ$ latitude grid averaged over PJ1 through PJ9, as presented by \citeA{20guillot_ammonia} using the same techniques as \citeA{17li}.  In order to fit the higher-than-expected microwave brightnesses measured by Juno \cite{17bolton}, these retrievals required NH$_3$ depletion compared to the standard JAMRT model, so our computed contribution functions generally probe higher pressures than those reported elsewhere in the literature \cite{17janssen}.  We assume a moist adiabat for the thermal structure based on NH$_3$, H$_2$S and H$_2$O, and all other atmospheric species and boundary conditions are as described in \citeA{20oyafuso}.  

The left-hand column of Fig. \ref{contfn} shows how the MWR channels probe higher altitudes with increasing emission angle, and how the contribution functions are relatively broad in the vertical direction.  The central column reveals how the latitudinal dependence derived by \citeA{20guillot_ammonia} influences the nadir contribution.  Because of the enhanced NH$_3$ retrieved in the Equatorial Zone, MWR channels tend to probe slightly higher in the equatorial region than they do in the neighbouring equatorial belts and the temperate mid-latitudes.  For the right column of Fig. \ref{contfn}, we identify the pressure at the peak of the contribution function for each emission angle for six scenarios:  three spatially averaged regions (northern mid-latitudes $20^\circ$N-$40^\circ$N, the equator $5^\circ$N-$5^\circ$S, and southern mid-latitudes $20^\circ$S-$40^\circ$S) and two different models of NH$_3$ opacity - those of \citeA{09hanley} and \citeA{16bellotti}.  As we are primarily concerned with mid-latitudes in this study, we average the mid-latitude contribution functions for both opacity models and both hemispheres, and employ a quadratic spline fit to interpolate over the emission angles in our experiments.  This provides smoothly varying functions for the angular dependence of the contribution functions at mid-latitudes, based on realistic NH$_3$ abundances. 

The calculations in Fig. \ref{contfn} reveal that, between emission angles of $0^\circ$ and $70^\circ$, MWR sounds a range of pressures in each channel:  1.4 cm (channel 6, 0.55-0.64 bar), 3.0 cm (channel 5, 0.8-1.6 bar), 5.75 cm (channel 4, 2.3-4.8 bar), 11.5 cm (channel 3, 6.0-13.8 bar), 24 cm (channel 2, 17.7-34.4 bar) and 50 cm (channel 1, 44-117 bar).  As expected, we find substantially less altitude sensitivity at the shortest wavelengths (channels 5 and 6, sounding $p<2$ bar) compared to the highest wavelengths (channels 1 and 2, sounding $p>20$ bar).  This is consistent with the extent of the limb darkening shown in Fig. \ref{limbdarken}.  We stress that the contribution functions remain extremely model dependent, varying with the retrieved ammonia abundances and assumptions about the lapse rate.  Furthermore, the peaks represent broad functions, with extensions to lower and higher pressures, particularly at the longest wavelengths \cite{17janssen}.  Channel 1 (50 cm) also displays significant sensitivity to pressures approaching 1000 bars, but this remains questionable given uncertainties about ammonia and water opacity at these long wavelengths \cite{20li_water}. 

Based on the contribution functions in Fig. \ref{contfn}, we can approximate the depth of the $\Delta_\mu$ transition from Fig. \ref{pearson}, where the flip from positive to negative correlations is observed in Channel 3 (11.5 cm).  In the northern temperate domain this occurs near $\theta=40-50^\circ$ (Fig. \ref{pearson}), placing the transition near 10-11 bars.  Similarly, the southern transition was at $\theta=70-80^\circ$, implying a transition nearer 4-6 bars.  These are averaged over all temperature latitudes in each hemisphere, and will be further refined below.

\subsection{Constructing a 2D Brightness Temperature Cross Section}

\begin{figure*}
\begin{center}
\includegraphics[angle=0,width=0.7\textwidth]{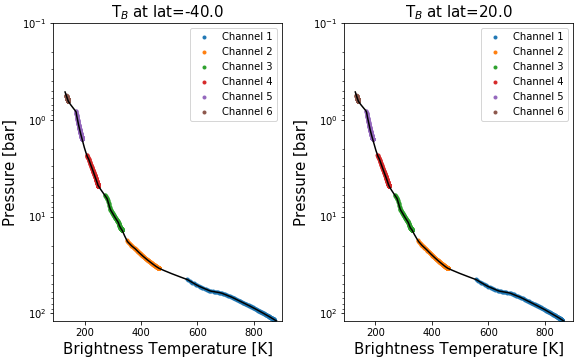}
\caption{Vertical profiles of $T_B$ at two different latitudes, estimated by assigning limb-darkened MWR measurements to discrete pressure levels using the contribution function peaks in Fig. \ref{contfn}. The y-axis indicates the pressure of the contribution peak at different emission angles, and different colours indicate different channels, with a smooth interpolation over regions without MWR sensitivity (retaining emission angles smaller than $70^\circ$).  Note that this is not from a spectral inversion, therefore does not represent kinetic temperatures - it is simply a reprojection of the MWR measurements.}
\label{TBprofiles}
\end{center}
\end{figure*}

We now use the emission-angle dependence of the MWR contribution functions (Fig. \ref{contfn}) to assign the model-independent $T_B(\phi,\mu)$ measurements from Fig. \ref{limbdarken} to a vertical pressure grid.  We stress that this is a method for reprojecting the $T_B$ measurements onto a pressure grid using a model-dependent contribution function, and should not be confused with a full inversion of the measurements to derive real kinetic temperatures.  This reprojection greatly expands the vertical sensitivity compared with the nadir-only approach, but we encounter substantial challenges, as shown in two example $T_B(p)$ profiles in Fig. \ref{TBprofiles}.  Firstly, the vertical sensitivity of adjacent MWR channels do not overlap with one another for emission angles smaller than $70^\circ$, so we are required to interpolate between them.  Secondly, adjacent channels do not line up sufficiently to produce a completely smooth vertical structure, resulting in some kinks in the $T_B(p)$ profiles.  This is particularly true for the transition between channels 5 and 6, where there is an offset of tens of degrees.  This is likely due to the assumptions underpinning the contribution function calculations: even though we have used realistic NH$_3$ distributions, differences in the NH$_3$ abundance could shift the peak sensitivity up and down and possibly allow better alignment of the channels.  Thirdly, we are effectively treating the contribution function as a delta function, assigning the $T_B$ to a unique pressure level and ignoring the broad range of pressures sounded in Fig. \ref{contfn} - this will be particularly problematic for channel 1, which has a broad contribution function reaching pressures of 1000 bars or greater.  And finally, the $T_B(\phi,\mu)$ has some dependence on the chosen functional form for the limb darkening (Eq. \ref{eq:coeff}) for high emission angles ($\mu<0.6$).

\begin{figure*}
\begin{center}
\includegraphics[angle=0,width=0.9\textwidth]{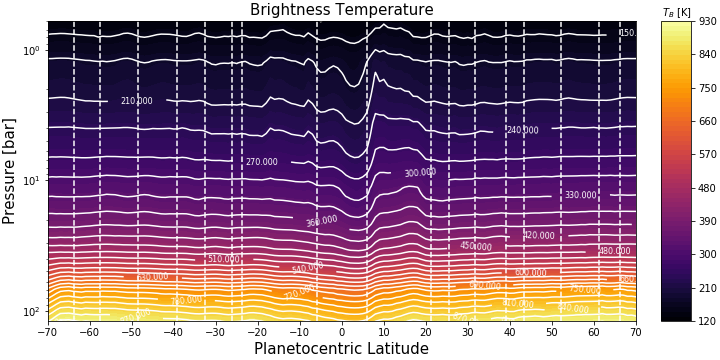}
\caption{2D cross-section of MWR brightness temperature $T_B(\phi,p)$, reprojected by assigning limb-darkened $T_B$ measurements to discrete pressure levels using the angular dependence of the contribution functions from Fig. \ref{contfn}.  Vertical dashed lines indicate the locations of the cloud-top prograde jets.}
\label{TBxsection}
\end{center}
\end{figure*}

\begin{figure*}
\begin{center}
\includegraphics[angle=0,width=1.0\textwidth]{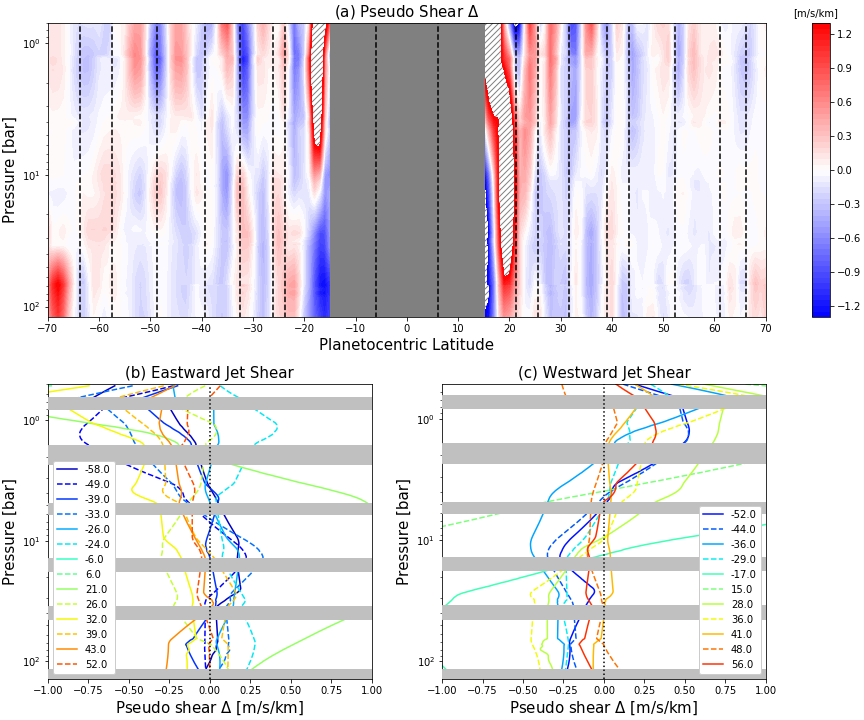}
\caption{(a) 2D cross-section of MWR brightness gradient $\Delta_\mu(\phi,p)$, or pseudo shear, in units of m/s/km, constructed from the $T_B(\phi,p)$ cross-section in Fig. \ref{TBxsection}.  The colour scale is saturated at $\pm1.3$ m/s/km to emphasise gradients at mid-latitudes, values of $\Delta_\mu$ exceeding this range are shown as grey hatches.  Tropical regions at latitudes less than $15^\circ$ are omitted. Vertical dashed lines indicate the locations of the cloud-top prograde jets. (b, c) Extracting the MWR pseudoshear $\Delta_\mu$ from (a) near to the locations of the eastward (b) and westward (c) jets, as shown by the planetocentric latitudes in the legends.  Grey horizontal bars indicate regions without MWR vertical sensitivity (as defined by Fig. \ref{contfn}) and discontinuities in the calculation of $\Delta_\mu$.  Tropical pseudoshears exceed $\pm1$ m/s/km over much of the domain, so cannot be seen on this figure.  The pseudoshear generally reverses sign near the 10-bar level, especially for southern-hemisphere jets.}
\label{gamma_xsection}
\end{center}
\end{figure*}

We construct $T_B(p)$ profiles for all latitudes and assemble them into a $T_B(\phi,p)$ cross section in Fig. \ref{TBxsection}, compared to the locations of the cloud-top zonal winds.  Although this has the appearance of a kinetic temperature cross section common in atmospheric physics, we caution that these $T_B$ values are the product of both temperature and opacity variations.  As for the nadir $T_B$ profiles in Fig. \ref{nadirTB}, the gradients away from the tropics are rather subtle, so we compute the `pseudo-shear' $\Delta_\mu$ for every pressure level in Fig. \ref{gamma_xsection}a.  Here, the transition from $\Delta_\mu>0$ (red) to $\Delta_\mu<0$ (blue), or vice versa, is visible throughout the temperate mid-latitudes (as well as the retrograde jets on the poleward edges of the NEB and SEB, discussed in Section \ref{nadirbrightness}).  

The transition occurs where $\Delta_\mu=0$ and is evidently latitude-dependent, so we plot $\Delta_\mu$ for individual eastward and westward jets in Fig. \ref{gamma_xsection}b-c, highlighting the high degree of variability from jet to jet.  The vertical trends in $\Delta_\mu$ are clearest for the broad westward jets, where Fig. \ref{gamma_xsection}c confirms that shears are generally positive for $p<10$ bars and negative for $p>10$ bars, although there is significant variability across the latitudes.  However, for the eastward jets the picture is unclear -  these are generally (but not always) experiencing negative $\Delta_\mu$ for $p<10$ bars, and they have small values ($\Delta_\mu<\pm0.25$ m/s/km) for $p>10$ bars, sometimes positive, sometimes negative.  We show in Section \ref{winds} that this weak $\Delta_\mu$, if interpreted as real kinetic temperature contrasts, might imply that eastward jets largely remain eastward at all depths to 100 bars, whereas the westward jets with larger $\Delta_\mu$ variations can change direction with depth.  The lack of clarity in $\Delta_\mu$ at the prograde jet locations could be a spatial-resolution effect related to their narrow or `sharp' latitudinal widths, compared to the broad retrograde jets.  Fig. \ref{gamma_xsection} suggests that the transition typically occurs in the 5-10 bar range, and is certainly easier to see in the locations of the westward jets.  In the next section, we explore what these pseudo-shears might imply about the zonal winds.

\subsection{Zonal Wind Interpretation}
\label{winds}

\subsubsection{Dry Thermal Wind Balance}
\label{windsT}

Prior to this point, we have been careful to describe the microwave brightness contrasts in terms of a pseudo-shear, $\Delta$, because both opacity variations (mainly NH$_3$) and kinetic temperature variations ($T$) could be responsible for gradients in $T_B$.  We now consider the extreme case where our measured $\Delta_\mu$ is assumed to be the true vertical windshear (i.e., that $T_B=T$, and that all brightness variations are considered to be due to kinetic temperature), and employ the `dry' thermal wind equation \cite{04holton}, neglecting contributions from molecular weight gradients (see Section \ref{windsNH3}):
\begin{eqnarray}
  \frac{\partial u}{\partial z}  \approx -\frac{g}{fT}\left(\frac{\partial T}{\partial y}\right)_p
  \label{eq:dryTWE}
\end{eqnarray}
Here $y$ is the north-south distance in kilometres, and the temperature gradients are measured on constant-pressure surfaces. We estimate the gravitational acceleration $g(p,\phi)$ using the combined gravitational and centrifugal potential of \citeA{20buccino}, reproducing their effective gravity at 1 bar.  We then use the ideal gas law to estimate the height $z(p,\phi)$, which reproduces the altitudes recorded by the Galileo probe \cite{98seiff}.  Both grids are provided in \ref{appS4}.

\begin{figure*}
\begin{center}
\includegraphics[angle=0,width=1.0\textwidth]{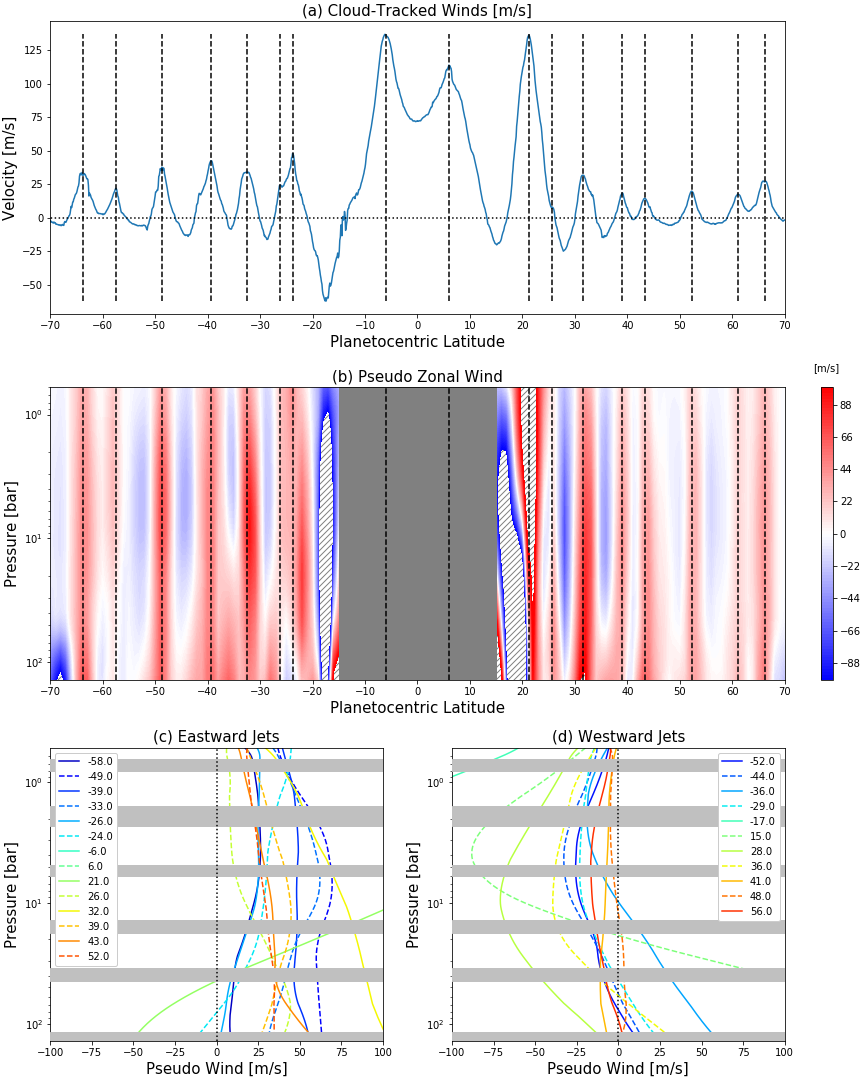}
\caption{Calculated pseudo winds (b) assuming that $\Delta_\mu$ can be equated to the vertical shear on the zonal winds (i.e., that $T=T_B$).  Integration is along the local vertical, rather than along cylinders parallel to the rotation axis.  Cloud-tracked winds from Cassini \cite{03porco}) are shown in panel (a) for comparison.  Speeds exceeding 100 m/s have been omitted (grey hatches), and speeds peak where $\Delta_\mu$ changes sign. Vertical dashed lines indicate the locations of the cloud-top prograde jets.  Low latitudes near the equator are omitted as the Coriolis parameter tends to zero (it varies as the sine of the latitude) and $\Delta_\mu$ therefore tends to infinity.  The lower panels show the MWR pseudowinds from (b), extracted near to the locations of the eastward (c) and westward (d) jets, as shown by the planetocentric latitudes in the legends.  Grey horizontal bars indicate regions without MWR vertical sensitivity (as defined by Fig. \ref{contfn}) and discontinuities in the calculation of $\Delta_\mu$.  Tropical windspeeds calculated in this manner exceed $\pm100$ m/s over much of the domain, so cannot be seen on this figure.  Note that this figure implies strengthening winds at $p>100$ bar, whereas Juno gravity measurements require that they must ultimately begin to decay at higher pressures \cite{18kaspi}.}
\label{winds_xsection}
\end{center}
\end{figure*}

We use Eq. \ref{eq:dryTWE} to integrate the cloud-top winds \cite{03porco} as a function of depth.  This quantity, the `pseudo-wind,' is shown as a cross-section in Fig. \ref{winds_xsection}b and for the individual jet locations in Fig. \ref{winds_xsection}c-d.  For simplicity, we integrate along the local vertical, rather than along cylinders parallel to the rotation axis, meaning that we cannot estimate winds close to the equator where the Coriolis parameter tends to zero.  However, as we are dealing here with a relatively shallow layer of atmosphere, with a small aspect ratio between the vertical and horizontal scales, this form of thermal wind is sufficient \cite{09kaspi}.  The latitude and depth-dependence of the gravity field is taken into account.  

For the mid-latitudes, Fig. \ref{winds_xsection} reveals the consequence of having a windshear that changes sign in the 5-14 bar region:  winds will increase with depth below the top-most clouds to reach an extremum in the 5-14 bar range, then the sense of the shear reverses to cause a decay with increasing depth.  For the prograde jets, the windshear is sufficiently weak that the jets mostly remain eastward throughout the domain sensed by MWR (i.e., $p<100$ bars) - most temperate jets at 100 bar would be in the 10-75 m/s range, not dissimilar from the speeds of those eastward jets at 1 bar.  The pseudo-shear is stronger for the retrograde jets, suggesting that the direction of the temperate jets could even switch from retrograde to prograde at pressures exceeding 20-30 bars (Fig. \ref{winds_xsection}d).  In most cases, the magnitude of these jets at 100 bars remains small ($<25$ m/s), although some of the jets approach 100 m/s at 100 bar, which is inconsistent with constraints imposed by the gravity measurements \cite{21galanti}.  This suggests that we cannot consider the $T_B$ variations in the deepest MWR channels to be solely driven by kinetic temperatures, and NH$_3$ (and potentially H$_2$O) must play a role.  Furthermore, we caution that the contribution functions for the MWR channels are highly model dependent, meaning that different assumptions about ammonia and water opacity could affect how the pseudo-shear $\Delta_\mu$ is distributed with height.  We also stress that integration of the windshear is prone to magnification of small errors with increasing depths, such that these deep winds should be treated with suspicion even if the assumption of $T_B=T$ were appropriate.

\subsubsection{Comparison to Juno Gravity}
\label{comparegrav}
It is natural to ask whether the inferred pseudo-winds are consistent with the results of Juno's gravity measurements \cite{18kaspi,18guillot}, which suggest a variety of potential wind profiles decaying to the 3000-km level, depending on the sensitivity to the measured odd gravity harmonics $J_3$, $J_5$, $J_7$ and $J_9$ \cite{20duer}. An increase in the temperate winds to the transition point at 5-14 bar, followed by a weak decay of the winds to higher pressures, is broadly consistent with the need for some form of decay profile in the interior \cite{18kaspi, 20kaspi}.  The gravity measurements are not directly sensitive to the winds at the altitudes sensed by MWR, but the analysis of the gravity data must assume a vertical profile for the velocity, which happens to be well matched to the cloud-top winds \cite{18kaspi}.  Indeed, \citeA{20duer} found that interior wind profiles that diverged from those measured at the cloud tops (i.e., depth-dependent flow profiles) could also be consistent with the gravity data, but concluded that they were statistically unlikely.  

The primary asymmetry in Jupiter's zonal winds is between the fastest retrograde jet in the south (the SEBs at $17.4^\circ$S) and the fastest prograde jet in the north (the NTBs at $21.3^\circ$N).  Fig. \ref{winds_xsection}b implies that this low-latitude asymmetry weakens with depth, suggesting our kinetic-temperature-only assumption (i.e., that $T_B=T$), and the implied strong shears on the equatorial jets in the $p>10$ bar region of Fig. \ref{winds_xsection}c, are not realistic.  Conversely, provided this low-latitude asymmetry is maintained, then the gravity measurements display a limited sensitivity to what the jets are doing at mid-latitudes poleward of $\pm25^\circ$, in terms of both direction and magnitude.  By retaining the observed cloud-top low-latitude winds within the $25^\circ$S to $25^\circ$N range, and introducing random velocity profiles for the temperate jets at higher latitudes, \citeA{21galanti} showed that this change has a limited effect on the goodness-of-fit to the odd gravity harmonics, as well as the even harmonics $J_6$, $J_8$, and $J_{10}$ (their Section 4 and Fig. 4).   In essence, a modification of the mid-latitude zonal jets below the clouds is not ruled out by the gravity data, provided that their magnitude remains small, which is the case in Fig. \ref{winds_xsection} with our extreme assumption that $\Delta_\mu$ represents the true vertical windshear.  Nevertheless, an optimal match to the gravity data still requires that the wind profile in the range $50^\circ$S to $50^\circ$N is unchanged from those measured at the cloud tops \cite{21galanti}.  It is more likely that both $T$ and NH$_3$ control the microwave brightness, such that the true vertical windshear is smaller than presented in Fig. \ref{gamma_xsection}, making it more consistent with the Juno gravity results.


\subsubsection{Comparison to Galileo Probe}
\label{galileo}
We can also compare the inferred structure of the pseudo winds from MWR to the only \textit{in situ} measurement of winds by the Galileo probe in 1995 \cite{98atkinson}.  The comparison is made complicated because (i) the probe descended into an anomalous tropospheric feature called a `5-$\mu$m hot spot' which may have influenced the measured winds, and (ii) this region was at the boundary between the EZ and NEB where the strongest $\Delta$ is measured \cite<related to the equatorial NH$_3$ enhancement,>[]{17li}. Nevertheless, the wind profile was found to approximately double from the 1-bar level to $\sim5$ bars, then level off and potentially show a weak decay with increasing pressure.  This was supported by Cassini cloud-tracking \cite{06li}, which suggested that the NEBs jet at $6^\circ$N strengthened with depth from the 0.5-bar level to the $\sim5$ bar level by more than 90 m/s, and also by an investigation of the stability of the zonal jets \cite{95dowling_SL9}, as discussed in Section \ref{shearstability}.  A decay of the zonal winds for $p<1$ bar is also supported by thermal-infrared observations \cite<e.g.,>[]{81pirraglia, 06simon, 16fletcher_texes}, suggesting that this shear region may actually extend from 0.5 to 5.0 bars.  

By taking gradients of the results from Galileo's Doppler Wind Experiment \cite{98atkinson}, we find that this is consistent with having negative vertical windshear for $p<5$ bars (approximately -2 m/s/km at 2 bars), and weakly positive windshear for $p>5$ bars (approximately 0.25 m/s/km at 10 bars).  The uncertainties on the Galileo wind profile start to grow large for $p>15$ bar, implying that both positive, zero, or negative windshears are possible \cite{01atkinson}.  Specifically for the NEB, this is inconsistent with the $\Delta$ measured by MWR (which remains negative throughout the 1-100 bar domain, presumably as a result of strong NH$_3$ contrasts such that the $T=T_B$ assumption is invalid here).  However, the Galileo-measured equatorial windshears are comparable in magnitude to the $\Delta$ in Fig. \ref{gamma_xsection} for mid-latitudes, suggesting that temperate jets that increase in strength down to the transition point, and then decay slowly with depth at higher pressures, are consistent with the structure observed by the Galileo probe, whether or not that measurement was truly representative of the equatorial zonal winds.  

Finally, \citeA{21galanti} explore whether Juno gravity measurements can still be reproduced if the zonal winds truly experience this doubling in strength from the cloud level to the 5-bar level, finding that plausible solutions can still be found, only with the winds decaying with a more baroclinic vertical profile compared to the \citeA{18kaspi} profile in the upper 2000 km, below which the winds decay more slowly, reaching 10\% of their original value at 3000 km.  This different wind decay could be considered as a viable alternative to the decay profiles in \citeA{18kaspi}, but additional constraints on the wind profiles in the 1-10 bar range are sorely needed, as discussed in Section \ref{shearstability}.

\subsubsection{Moist Thermal Wind Balance}
\label{windsNH3}

In this Section we describe how latitudinal gradients in molecular weight can still lead to vertical windshear, even if the kinetic temperature remains uniform.  In the case where both compositional and thermal variations result in latitudinal density gradients \textit{along constant-pressure surfaces}, we express the geostrophic thermal wind equation \cite{04holton} in its less familiar `moist' or `virtual' form \cite<sometimes known as a `humidity wind' equation,>[]{91sun} in altitude coordinates $z$:
\begin{eqnarray}
  \frac{fT}{g}\frac{\partial u}{\partial z} = -\left(\frac{\partial T_v}{\partial y}\right)_p
\end{eqnarray}
where symbols have the same meanings as in Section \ref{windsT}.  \citeA{91sun} demonstrated that compositional gradients could have a significant influence on the windshear in hydrogen-rich atmospheres, most important with the observed enrichments of Uranus and Neptune over solar composition, but here we explore the implications for Jupiter's troposphere.  The virtual temperature $T_v$ is defined as:
\begin{eqnarray}
  T_v=\frac{T}{1+\Sigma \alpha_c q_c}
\end{eqnarray}
Here $q_c$ is the mole fraction, $\alpha_c$ is a coefficient for each constituent equal to $(\mu_c/\mu_d)-1$, the ratio of the molecular weight of the constituent ($\mu_c$) to the molecular weight of dry air ($\mu_d$).  The $\Sigma$ symbol implies a sum over the relevant gases (NH$_3$, H$_2$S, H$_2$O).  We do not directly relate $T_v$ to the observed $T_B$ gradients, but introduce it simply to account for the effects of molecular weight gradients on vertical shears.  The derivation below differs from Eq. 7 of \citeA{91sun} because we use mole fractions, whereas they used mass mixing ratios.  In the case where these constituents are considered to be variable, we adjust the thermal wind equation to become:
\begin{eqnarray}
  \frac{fT}{g}\frac{\partial u}{\partial z} = -\frac{\partial }{\partial y}\left(\frac{T}{1+\Sigma \alpha_{c}q_{c}}\right) \\
  = -\frac{1}{1+\Sigma \alpha_{c}q_{c}}\left(\frac{\partial T}{\partial y} - \frac{T}{1+\Sigma \alpha_{c}q_{c}}\frac{\partial}{\partial y}(\Sigma \alpha_{c}q_{c})\right)
  \label{eq:fullTWE}
\end{eqnarray}
If we retain the molecular weight contributions of all three condensables, but assume that both H$_2$S and H$_2$O are latitudinally uniform to remove their derivatives, then we can rewrite the $T_v$ gradient considering only contributions from the temperature and NH$_3$ gradients:
\begin{eqnarray}
  \frac{fT}{g}\frac{\partial u}{\partial z}  = -\frac{1}{1+\Sigma \alpha_{c}q_{c}}\left(\frac{\partial T}{\partial y} - \frac{T\alpha_{NH_3}}{1+\Sigma \alpha_{c}q_{c}}\frac{\partial q_{NH_3}}{\partial y}\right)
\end{eqnarray}

In the case where we assume no latitudinal ammonia gradients, and with $\Sigma \alpha_{c}q_{c}<<1$ (a reasonable assumption in the upper troposphere where mole fractions of each species are $<10^{-3}$, but more questionable at depth), this simplifies to the familiar dry thermal wind equation in Eq. \ref{eq:dryTWE}, as discussed in Section \ref{windsT}.  However, if we assume negligible latitudinal contrasts in temperature, following previous MWR analyses \cite{17li, 17ingersoll}, and again assuming $\Sigma \alpha_{c}q_{c}<<1$, then we find that ammonia gradients can still result in vertical windshear:
\begin{eqnarray}
  \frac{\partial u}{\partial z}  \approx + \frac{g\alpha_{NH_3}}{f}\frac{\partial q_{NH_3}}{\partial y}
  \label{eq:wetTWE}
\end{eqnarray}
Here $\alpha_{NH_3}=(\mu_{NH_3}/\mu_d)-1=6.36$, with $\mu_{NH_3}=17.031$ g/mol and the dry molecular weight of jovian air is $\mu_d\approx2.313$ g/mol, assuming 86.26\% H$_2$, 13.54\% He, and 0.20\% CH$_4$ \cite{98vonzahn,04wong_gal}.  Note the change in sign between the two forms of the wind equation (Eq. \ref{eq:dryTWE} and \ref{eq:wetTWE}), and how it relates to the MWR brightness temperature observations.  Local maxima in microwave brightness over belts in the upper troposphere ($p<5$ bar) would still be in balance with negative $\partial u/\partial z$ (i.e., wind decay with height) irrespective of whether this is due to an \textit{increased} temperature or an NH$_3$ minimum.  Local minima in $T_B$ in the deeper troposphere ($p>10$ bar) would still be in balance with positive $\partial u/\partial z$ (i.e., wind decay with depth), irrespective of whether this is due to an \textit{decreased} temperature or an NH$_3$ maximum.  In both the temperature-only and the ammonia-only cases, the vertical windshear would have the same sign.  But how significant is this effect?  

\begin{figure*}
\begin{center}
\includegraphics[angle=0,width=1.0\textwidth]{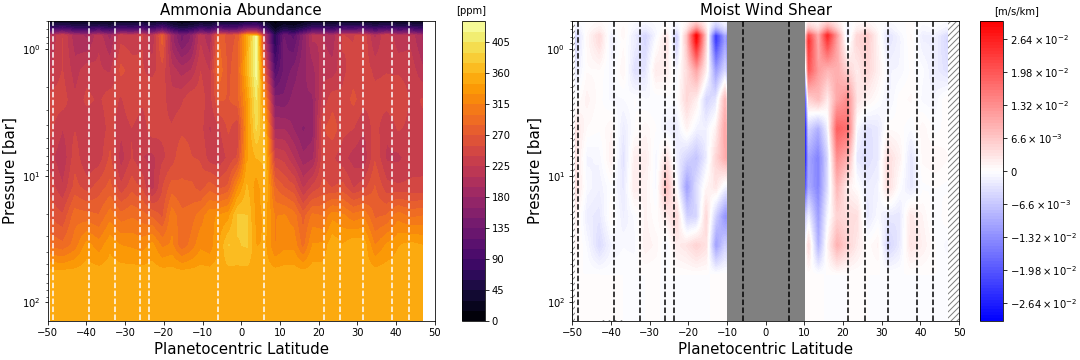}
\caption{Zonal-mean cross section of ammonia derived by \citeA{20guillot_ammonia} based on the technique of \citeA{17li}.  The gradients are used to estimate the moist shear based on NH$_3$ alone, which is some $50\times$ smaller than that in Fig. \ref{gamma_xsection} for mid-latitudes.}
\label{moist_shear}
\end{center}
\end{figure*}

\citeA{20guillot_ammonia} provide a retrieved latitude cross-section of NH$_3$ abundances averaged over PJ1 to PJ9 which we can use to measure $\partial q_{\rm{NH_3}}/\partial y$ as an estimate of $\partial u/\partial z$ (Fig. \ref{moist_shear}). Although the resolution of their inversion is lower than the resolution of the MWR brightness temperature used in this study, Fig. \ref{moist_shear} confirms the flip in sign of the shear as a function of depth, and shows that the peaks in the shear remain co-located with the locations of Jupiter's cloud-top jets.  Note that this NH$_3$ cross-section was the basis for our contribution function calculation in Fig. \ref{contfn}.

Based on NH$_3$ alone, the shear is strongest near the equator, approaching -0.25 m/s/km for the NEBs jet (not shown) in the 0.6-2.0 bar region, which is approximately 10\% of the shear needed to explain those measured by the Galileo probe.  In the temperate mid-latitudes, we find $\partial q_{\rm{NH_3}}/\partial y$ in the range $\pm1.5\times10^{-8}$ km$^{-1}$, which equates to windshears in the range $\pm0.03$ m/s/km, at least $50\times$ smaller than the brightness-temperature derived $\Delta_\mu$ in Fig. \ref{gamma_xsection}(a).  On this basis, if NH$_3$ contrasts are the only significant contributor to MWR brightness gradients, then the integrated mid-latitude winds will be largely barotropic in the 1-100 bar range. 

As a final thought experiment, we extended Eq. \ref{eq:wetTWE} to include the influence of H$_2$O, still assuming that $\Sigma(\alpha_c q_c)<<1$:
\begin{eqnarray}
  \frac{\partial u}{\partial z}  \approx + \frac{g}{f} \left( \alpha_{NH_3}\frac{\partial q_{NH_3}}{\partial y} +  \alpha_{H_2O}\frac{\partial q_{H_2O}}{\partial y}\right)
  \label{eq:wetTWE_h2o}
\end{eqnarray}
Here $\alpha_{H_2O}=(\mu_{H_2O}/\mu_d)-1=6.78$, with $\mu_{H_2O}=18.015$ g/mol.  The latitudinal distribution of H$_2$O is currently unknown, so we estimate $\partial q_{\rm{H_2O}}/\partial y$ by scaling the equatorial water profile of \citeA{20li_water} using the latitude dependence of the NH$_3$ results in Fig. \ref{moist_shear}.  This is a very crude assumption, but supposes that the same processes shaping the NH$_3$ distribution (Ferrel cells or precipitation, see Section \ref{discuss}) are also governing the as-yet-unmeasured H$_2$O distribution \cite{20guillot_ammonia}.  The contribution of water to moist thermal wind balance is approximately $3\times$ larger than that of ammonia - at mid-latitudes, in the 5-50 bar region, this would produce shears of $\pm0.1$ m/s/km (a factor of $\sim10$ smaller than those shown at mid-latitudes in Fig. \ref{gamma_xsection}), rising to -1 m/s/km for the NEBs jet, which is too large (and too negative) to be consistent with the windshear directly measured by the Galileo Probe for $p>5$ bar, potentially suggesting that such strong water contrasts are unlikely in the equatorial domain.  

The effect of such a weak moist windshear at mid-latitudes would be that the winds would be almost barotropic over the domain sounded by MWR (1-100 bars), which would also be consistent with the Juno gravity measurements \cite{21galanti}.  However, it is counter to that shown from the dry windshear equation in Fig. \ref{winds_xsection}, and counter to the Galileo probe wind measurements that showed strong variability with depth.  There remains much debate over whether the winds observed by Galileo \cite{98atkinson, 06li_wind} were a local consequence of the Rossby-wave dynamics of the 5-$\mu$m hot spot \cite{00showman}, or globally representative of the shear on the NEBs jet.  If the latter is true, then the Galileo winds suggest the need for some kinetic temperature contrasts (i.e., dry windshear) in at least the 0.5-5.0 bar region sounded by MWR channels 4-6, because the moist windshears discussed above are insufficient.  However, without being able to uniquely separate ammonia and kinetic temperatures in a microwave inversion, MWR conclusions about zonal winds still range from nearly \textit{vertically uniform} to \textit{vertically variable} with a transition near 5-14 bars, and it might even be possible that the dry and moist windshears actually oppose one another at some locations (i.e., a region that is both warm and enriched in volatiles).  Additional constraints on deep kinetic temperatures are sorely needed, as we explore in the next section.

\subsubsection{Deep Thermal Contrasts}
\label{shearstability}

Breaking the degeneracy between deep temperature and ammonia contrasts via remote sensing alone (e.g., microwave and infrared) remains a challenge.  However, we can gain insights on the likelihood of deep temperature gradients (and winds that increase in speed from the cloud tops to the 5-10-bar level) by (i) considering the stability of the zonal wind solutions; and (ii) exploring the results of deep convection models.

For the former, the top-down constraint on the jet structure offered by vorticity measurements support the suggestion that the winds must increase with depth from the cloud-tops to regions near the water cloud \cite{95dowling_SL9}.  As the meridional gradient of the potential vorticity changes sign at multiple locations \cite<e.g.,>[]{06read_jup}, the cloud-top winds (and our inferred winds at depth) have multiple critical latitudes which could be stable, unstable, or neutrally stable \cite{95dowling_SL9, 20dowling}.  Before the descent of the Galileo probe, \citeA{95dowling_SL9} used Voyager-era vorticity measurements \cite{86limaye} and a shear-stability analysis to determine Jupiter’s deep wind profile in the 5-8 bar region.  To make the cloud-top critical latitudes stable, rather than marginally stable, required an increase in the amplitude of the underlying eastward jets compared to the cloud-top jets by a factor of approximately two, with larger changes at lower latitudes than at mid-latitudes.  The magnitude of the change depended on the first-baroclinic deformation length, $L_d$, which remains rather uncertain at depth.  Their suggested negative vertical shear of the zonal winds between the tropopause and the 5-8 bar level was later shown to be consistent with Galileo probe results \cite{98atkinson}, and qualitatively supports our suggestion that winds strengthen between the cloud-tops and the jovicline in the upper cell (i.e., that kinetic temperatures must vary with latitude, helping to explain the negative pseudoshear in shallow-sounding MWR channels 4-6).

 
Finally, deep-shell models of turbulent convection in rapidly-rotating fluid planets produce nested cylindrical flows aligned with the rotation axis, with alternating zonal jet structures and associated meridional temperature contrasts \cite{08aurnou, 16heimpel}.  These models produce axial thermal plumes parallel to the rotation axis, with the jets acting as barriers to cylindrically radial heat transfer.  With warm fluid on the equatorward sides of jets, and cool fluid on the poleward side, the model of \citeA{08aurnou} exhibits a pattern qualitatively similar to our deep circulation cells ($p>10$ bars) and opposite to those above the jovicline ($p<10$ bars).  The axial wind structures appropriate for the deeper layers still needs to be properly connected to the radial wind structures in the shallow layers observed by MWR, but this is a compelling connection suggesting that deep kinetic temperature perturbations (and associated windshear) cannot be ruled out as contributing to the MWR contrasts in the 1-100 bar region.

\section{Discussion}
\label{discuss}

Juno MWR observations between August 2016 and April 2018 have revealed that mid-latitude gradients in microwave brightness ($\Delta$) are well correlated with the locations of the cloud-top zonal winds, and that this correlation shifts from being negative in shallow-sounding channels (4-6, approximately $p<5$ bars) to positive in deep-sounding channels (1-3, approximately $p>5$ bars).  As a consequence, cyclonic belts that appear microwave-bright at shallow pressures (i.e., depleted in volatiles and/or physically warm) become microwave dark at higher pressures in the deep atmosphere (i.e., enriched in volatiles and/or physically cool).  Using the dependence of $\Delta_\mu$ on emission angle, and a model-dependent estimate of the MWR contribution functions for each wavelength and viewing geometry, we find that this transition pressure varies considerably with latitude, but is typically found in the 5-10 bar region.  The transition is clearest in the southern hemisphere where correlation coefficients are larger, but is also visible in the northern hemisphere.  The transition is easier to discern for the broad retrograde jets than the narrow prograde jets, but this may be a consequence of the spatial resolution of MWR failing to capture gradients over narrow (i.e., $1^\circ$) latitude ranges.

The belts and zones therefore change their character as a function of depth, irrespective of how the microwave spectra are interpreted (e.g., as compositional variations, temperature variations, or a combination of both).  This had been previously noted by \citeA{17ingersoll} based solely on the PJ1 (August 2016) observations, but they had suggested that the relationship between temperate brightness gradients and the zonal jets was rather poor.  Using these same PJ1 data, \citeA{20duer} also showed the correlation between winds and MWR brightness observations.  Using data from subsequent perijoves, filtering via the deconvolution process of \citeA{20oyafuso}, and by taking the gradient $\Delta$, we have shown that the correlation with the cloud-top winds is actually much better than originally thought.  

We now explore the potential consequences of this transition, which we call the `jovicline' via analogy to the thermocline in Earth's oceans (the transition layer between warm waters near the surface and cool waters at depth) or the tachocline in the Sun's interior (the transition layer between the interior radiative zone and upper convective zone).  However, whereas the terrestrial thermocline is a region with a sharp change in vertical temperature gradient, and resulting change from low-density surface waters to high-density deep waters (the pycnocline), the jovicline is a transitional level where Jupiter's belt-zone contrasts, and hence the vertical shears, appear to change sign.  The jovicline is not to be confused with Jupiter's `planetary tachocline' at much higher pressures, where Ohmic dissipation on the flows becomes important \cite{11heimpel}.  To our knowledge, the first use of the word `thermocline' in a description of Jupiter’s atmosphere appeared in Arthur C. Clarke’s science fiction story, ``A Meeting With Medusa,'' during the voyage of the \textit{Kon Tiki} balloon down into the cloud layers of Jupiter \cite{72clarke}.  Earth's oceanographic `clines' serve as a barrier to vertical mixing, separating the circulations of the shallow and deep layers.  \textit{Might it be possible for the jovicline to act as a similar barrier?}

\subsection{Stacked Meridional Circulation Cells}

\begin{figure*}
\begin{center}
\includegraphics[angle=0,width=1.1\textwidth]{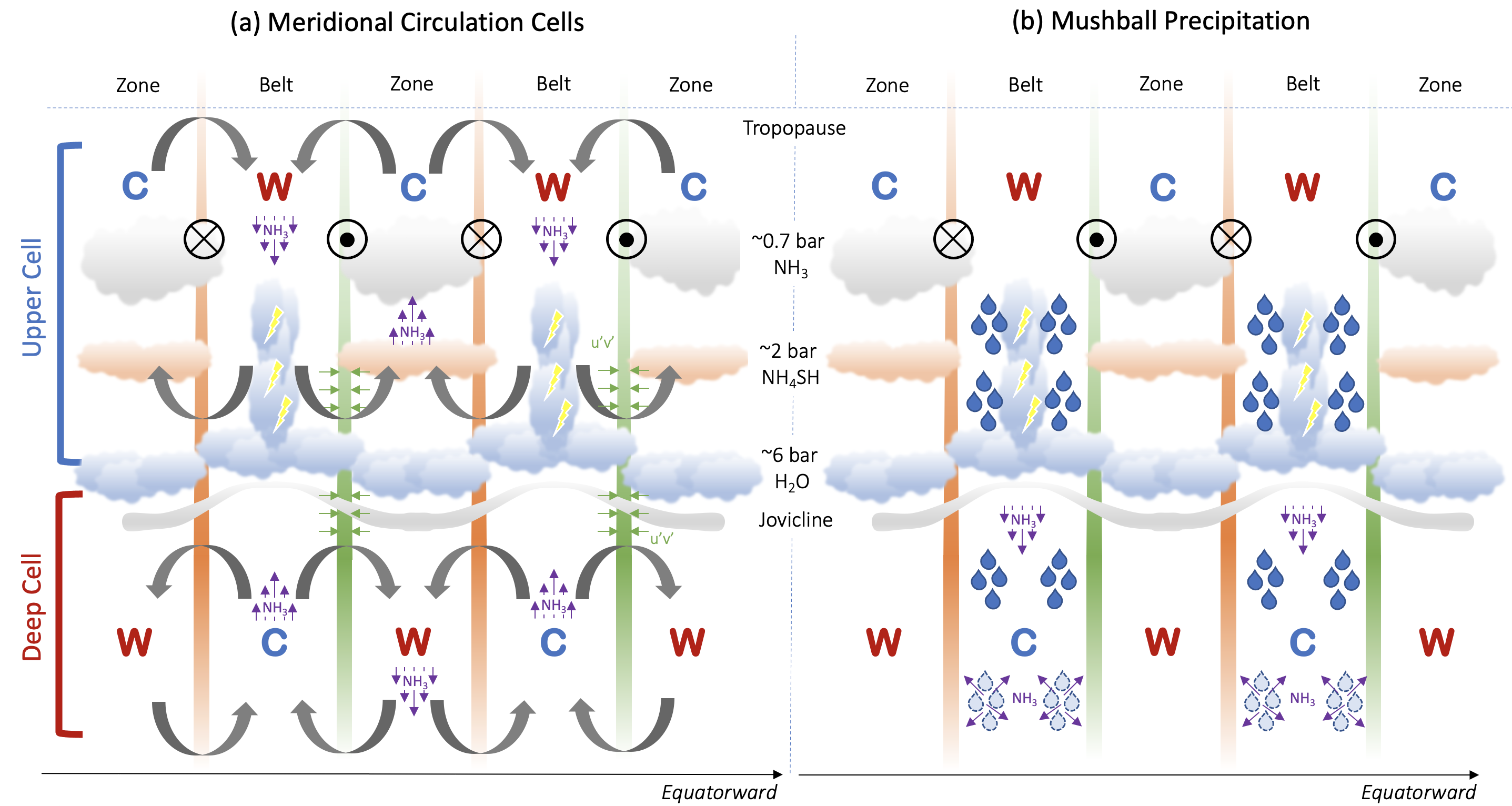}
\caption{Conceptual diagrams of (a) the stacked system of meridional cells \cite<adapted from>[]{05showman,20fletcher_beltzone}; and (b) mushball precipitation \cite{20guillot_ammonia}.  We stress that reality is likely to combine both of these concepts, and all altitudes are qualitative.  In both diagrams, high microwave brightness is denoted by a red `W' (warm), low microwave brightness is denoted by a blue `C' (cool); storm plumes are indicated as rising clouds with lightning flashes.  The equator is to the right, such that belts have prograde jets on their equatorward edges. Eastward prograde jets are green (with a circular dot indicating motion out of the page) with eddy-momentum flux convergence (small green arrows); westward retrograde jets are orange (with a circular cross indicating motion into the page).  The colouration of the green and orange bars indicate wind strengthening through the upper cell and wind decay with depth in the deep cell (`dry convective layer').  The jovicline is shown in grey, co-located with the stable stratification of the water cloud.  Purple arrows indicate general ammonia depletion or enrichment, either as a consequence of meridional circulation (grey curved arrows, left) or as a consequence of sequestration in `mushballs', precipitation, and re-evaporation at great depth (droplets, right), leading to steep vertical NH$_3$ gradients in the belts.}
\label{circulation}
\end{center}
\end{figure*}

As described in Section \ref{intro}, the concept of multiple tiers of stacked circulation cells \cite{00ingersoll, 05showman, 20fletcher_beltzone} has been used as a possible resolution to the discrepancy between (i) zone-to-belt transport and subsidence in belts above the clouds inferred from Jupiter's upper tropospheric temperatures and composition; and (ii) belt-to-zone transport in Ferrel-like cells below the clouds and upwelling in belts inferred from the prevalence of lightning in Jupiter's belts \cite{00ingersoll} and the meridional flow required to balance the eddy-momentum flux convergence on the prograde jets (Fig. \ref{circulation}a).  The change in the microwave brightness contrast across the transition would be consistent with NH$_3$ (and potentially other gaseous species) being locally depleted in belts in the upper tier, and locally enhanced in belts in the deeper tier \cite{05showman, 17ingersoll}.  The transition between these tiers was assumed to exist somewhere within the cloud-forming region \cite{05showman}, where vertical currents would meet and diverge \cite<e.g.,>[assumed it to be near the top-most condensate clouds]{20fletcher_beltzone}.  Furthermore, numerical simulations of giant planet tropospheres, and particularly the Ferrel-like circulations away from the equator \cite{05yamazaki, 18young, 20spiga}, do appear to support changes in meridional circulation as a function of height, possibly associated with a shift from eddy-forcing of zonal jets within the clouds \cite{06showman, 08lian, 10liu} to a domain of eddy dissipation and wind decay in the upper troposphere.

However, this study suggests that whilst a transition does exist, its likely location is deeper, at or below the water cloud as depicted in the cartoon in Fig. \ref{circulation}.  Equilibrium cloud condensation models \cite{99atreya} predict that Jupiter's primary volatiles (NH$_3$, H$_2$S and H$_2$O) will form cloud decks in the 0.7-to-7-bar range.  Specifically, in the absence of microphysical processes and precipitation, solar enrichment of Jupiter's elemental abundances would place the base of the water cloud near 5.7 bars, whereas a $3\times$solar enrichment would place it nearer 7.2 bars \cite{99atreya}.  Given that Jupiter's tropospheric composition is spatially variable \cite{86gierasch, 06achterberg, 16fletcher_texes, 16depater, 17li}, and that the $T(p)$ and lapse rate may differ between belts and zones, it is reasonable to assume that the water cloud base rises and falls (in the 5-8 bar range) depending on the properties of the atmospheric band.  Fig. \ref{gamma_xsection}a does imply that the transition varies with height on the scale of the belts and zones.  

The co-location of the predicted water cloud base with the jovicline may be no coincidence, in that this signifies the transition zone between the dissipative upper layer and the Ferrel-like circulations of the deeper troposphere.  The formation of the water cloud produces a density stratification \cite{14sugiyama, 15li, 16thomson}, whereby increased molecular weight of the water produces a stabilising layer that may serve to segregate the deeper circulations in the dry adiabatic layer from those of the moist upper cells.  This stable inversion layer can actually inhibit moist convection until potential energy has accumulated to some critical level, leading to the episodic convective outbursts that appear common within Jupiter's belts \cite{08sanchez, 17fletcher_seb, 17sanchez_NTB, 19depater_alma, 20wong}, maybe as part of a `charge-recharge' cycle of CAPE based on water.  Note that the upper tier above the water condensation altitude is sometimes referred to as the `weather layer', but given recent suggestions that NH$_3$ contrasts extend very deep \cite{17bolton, 17li}, we refrain from using this terminology.

In the stacked-cell hypothesis in Fig. \ref{circulation}a, belts in the upper cell would be regions of large-scale subsidence creating warm temperatures (and therefore an absence of condensed clouds), zonal wind strengthening with depth \cite{81pirraglia}, local ammonia depletion, and therefore a high microwave brightness as we see in the MWR observations for $p<5$ bar.  Conversely, belts in the deeper Ferrel-like cells would be regions of upwelling, with local ammonia enrichment and cooling in regions of adiabatic expansion (and therefore zonal wind decay with depth), leading to the microwave-dark belts that we see in the MWR observations for $p > 10$ bar.  Note that this discussion assumes an NH$_3$ abundance that \textit{decreases} with height throughout both upper and lower tiers, counter to the weak and currently unexplained \textit{increase} of NH$_3$ with height suggested by MWR inversions in the 2-6 bar region \cite{17li}.  As explored in Section \ref{winds}, the observed temperature and/or composition gradients could imply zonal winds increasing in strength from the tropopause to the jovicline, then decaying away slowly with increasing pressure into the dry adiabatic layers, although the strength of the windshear depends on whether temperature or abundance variations are responsible for the observed microwave brightness contrasts.  The observed cloud-top winds could therefore be an underestimate of the maximum windspeeds in the upper troposphere (Fig. \ref{winds_xsection}b).  

However, this contrived picture is incomplete - it does not explain the extreme ammonia enrichment at the equator, nor does it explain why the global-scale NH$_3$ depletion appears to extend to the 40-60 bar level \cite{17ingersoll, 17li}, far deeper than simple precipitation might suggest \cite<e.g., via the inclusion of ammonia rain,>[]{19li}.  Ferrel-like circulation cells below the jovicline \cite{05showman, 18young}, balancing eddy-momentum flux convergence on the prograde jets \cite{06salyk}, could extend deep even if the forcing is shallow \cite{08lian}, driving temperature and compositional variability at tens of bars.  The belt/zone meridional circulations inferred here may be superimposed onto this larger-scale structure (equatorial NH$_3$ enrichment, mid-latitude NH$_3$ depletion) driven by precipitation, to be explored in the next section.  Lightning could still be prevalent in the belts in Fig. \ref{circulation}a with this deeper jovicline, if rising motion from the deep `dry-convecting' layer provides the initial instability to initiate buoyant moist convection and lightning in the water-cloud layers and above \cite{89dowling, 16thomson}.  This could work if the stably-stratified transition zone were thinner (and easier to overcome) in the belts compared to the zones - a possible consequence of winds that decay with depth into the deeper layers \cite{16thomson}.  



\subsection{Precipitation and Microwave Brightness}

The complexity of the stacked-cells hypothesis may yet be its undoing, so we should ask \textit{whether vertical and meridional motions are truly required to explain the transition in the microwave belt/zone contrasts}.  Recent work by \citeA{20guillot_mushball} suggested that partially-melted hailstones of ammonia dissolved in water ice (nicknamed `mushballs') could form at 1-2 bar when water is lofted upwards during powerful storms \cite<this is also the level of shallow lightning flashes recently discovered by Juno,>[]{20becker}.  These mushballs then fall deep below the expected water cloud (Fig. \ref{circulation}b), to 5-30 bar depending on their properties and the available water ice, where they evaporate, causing cold and volatile-rich evaporative downdrafts that further deplete the condensates.  \citeA{20guillot_mushball} use this process to explain the observed deep depletion of NH$_3$ down to the 20-30 bar region \cite{17li, 17ingersoll}.  

As storms are more prevalent within Jupiter's belts, we might expect NH$_3$ depletion in the upper troposphere to be strongest here (producing the microwave-bright belts for $p<5$ bars).  Similarly, as the mushballs evaporate to relinquish their ammonia (and water), they increase the mean molecular weight in the deeper troposphere, and generate cool downdrafts \cite{14sugiyama}. This could lead to a localised NH$_3$ enhancement in the belts at depth (i.e., microwave-dark belts at $p>10$ bars).  Combined, this leads to a steep $dq_{\rm{NH_3}}/dz$ gradient in the belts, shown in Fig. \ref{circulation}b, as precipitation dominates over any upward mixing.  Conversely,  \citeA{20guillot_ammonia} suggested that the absence of storms and mushballs in the Equatorial Zone was responsible for the vertical homogeneity of the NH$_3$ distribution there.  Here we suggest that a shallow $dq_{NH_3}/dz$ gradient could also persist in the extratropical zones for the same reason (i.e., upward mixing dominates over precipitation), providing the contrast to the larger $dq_{NH_3}/dz$ in the stormy belts.  At high pressures, slow horizontal mixing would serve to transport NH$_3$ from belts into zones, and vice versa at lower pressures.  

\citeA{20guillot_ammonia} parameterised the storm frequency using the MWR observations of \citeA{18brown} - however, the detection of lightning sferics in the microwave still placed non-negligible storm flashes in regions considered as zones, and an imperfect relationship between local maxima in the storm rates and the location of the belts.  For this reason, the model of \citeA{20guillot_ammonia} (their Fig. 6) does not show the banded structure in the temperate domain that is observed in our study.  However, if the storm frequency were simply parameterised as being high in the belts and negligible in the zones, we might expect to recover the banding in Fig. \ref{limbdarken} from this mushball model.  In this scenario, the jovicline (and the base of the expected water cloud) is simply the level at which the abundances of NH$_3$ in the belts and zone are approximately equivalent (Fig. \ref{circulation}b), leading to $\Delta=0$ m/s/km.  

As with the stacked-cells hypothesis, the mushball hypothesis remains incomplete.  We still need some form of vertical/meridional circulation in the upper troposphere to explain the observed temperatures and distribution of disequilibrium species (e.g., PH$_3$ enhanced over zones and depleted over belts, and vice versa for para-H$_2$); and in the deeper troposphere to balance the eddy-momentum flux convergence into the prograde jets \cite<e.g., see review by>[]{20fletcher_beltzone}.  Given the density stratification contrasts associated with belt/zone differences in mushball formation and evaporation, we might expect some degree of secondary circulation and slow mixing that changes character with depth.  So it is possible that the observed transition in belt/zone properties can be explained by a combination of meridional Ferrel-like circulation and mushball precipitation, blending together the processes in Fig. \ref{circulation}.  Distinguishing between these scenarios may have to wait for more comprehensive general circulation models that include the mushball process, and we await such models with great interest.

\section{Conclusion}
\label{conclude}

Jupiter's temperate mid-latitudes (approximately $\pm20-60^\circ$ latitude) exhibit a banded structure in microwave brightness, characterised by the gradient $\Delta$ that is well correlated with the observed latitudes of the cloud-top zonal winds.  However, this correlation changes sign between Juno's shallow-sounding channels ($p\sim$0.6-5 bar, $\lambda=1.4-5.75$ cm) and deep-sounding channels ($p\sim$6-100 bars, $\lambda=11.5-50$ cm), implying that Jupiter's belts and zones change their character as a function of depth (Fig. \ref{circulation}). The identification of the transition is based on the MWR data alone, independent of radiative transfer and degenerate spectral inversions, but assigning a depth requires model-dependent calculations of microwave contribution functions as a function of emission angle.  Based on those calculations, we find that the transition between these two regimes (the `jovicline') appears to separate the layer above the water-condensation region (at 5-8 bars) from the deeper dry adiabatic troposphere.  The co-location of this transition with the base of the putative water cloud may be no coincidence, as the molecular weight gradient may have a stabilising influence, separating two regimes.  

If we interpret $\Delta_\mu$ as being a true reflection of the vertical wind shear (either weak shear associated with compositional gradients, or stronger shear associated with kinetic temperature gradients), then the gradients imply winds that strengthen from the cloud-tops to the jovicline, and then weaken at higher pressures. This is qualitatively consistent with \textit{in situ} winds measured by Galileo and with winds inferred from shear instability analyses, but we caution that (i) tropical contrasts are likely primarily related to ammonia \cite{17li}, and (ii) the strong hemispheric asymmetry between the retrograde SEBs and prograde NTBs jets \cite<e.g.,>[]{18kaspi, 20duer} must be maintained to match Juno's gravity measurements (Fig. \ref{winds_xsection}), such that the observed microwave contrasts at low latitudes cannot be solely driven by kinetic temperatures.  But at temperate latitudes polewards of $\pm25^\circ$, the location and direction of the extratropical jets have a smaller influence on the measured gravity field \cite{21galanti}, such that small wind variations with depth at mid-latitudes cannot be ruled out.  These results hint at the baroclinic nature of Jupiter's atmosphere both above and below the jovicline, but that the jovicline itself may be a region where horizontal temperatures and ammonia distribution are more uniform (leading to a barotropic region where shear tends to zero and winds are more uniform with height).

Using the signatures of gravity waves in the Doppler residuals from the Galileo probe, \citeA{01allison} explored the evidence for an increase in the static stability below the 5-bar level, suggesting a statically stable layer that they call the ``thermocline.''  This was supported by the idea that large-scale oscillations in thermal emission in the upper troposphere could be due to Rossby waves leaking out of a deeper waveguide \cite{90allison, 98ortiz}, and the inferences of a deep stable layer from the propagation of wavefronts from the Shoemaker-Levy 9 impact \cite{94ingersoll}.  Statically stable layers were also detected in data from the Galileo Probe Atmospheric Structure Investigation at 8 bar and 14 bar in the probe entry site \cite{98seiff, 02magalhaes}, coinciding with compositional gradients measured by the Galileo Probe Mass Spectrometer \cite{04wong_gal,09wong}.  This inferred deep stable layer could be related to the molecular static stability in the water cloud layer, stabilising the jovicline region.  

We explored potential explanations for why the microwave gradients flip sign above and below the jovicline.  Maybe stacked tiers of meridional circulation cells \cite{00ingersoll, 05showman, 20fletcher_beltzone} are the culprit, with belts exhibiting subsidence (NH$_3$ depletion and warming) above the jovicline and upwelling (NH$_3$ enhancement and local cooling) at higher pressures.  The Ferrel-like circulation of the deeper cell may be easier to explain because the eddy-momentum flux convergence has been observed \cite{06salyk} and modelled \cite{18young}.  Conversely, the circulation of the upper cell (where winds decay with altitude through the cloud layers) remains hard to explain because no drag force has yet been adequately identified, although the breaking of vertically-propagating waves remains a possible dissipation source \cite{86gierasch, 89pirraglia, 93orsolini}.  Maybe the latitudinal dependence of storms and precipitation, particularly in the properties of `mushballs' \cite{20guillot_mushball}, means that the vertical NH$_3$ gradient is steeper in the belts (lots of storms and associated precipitates) and shallower the zones (less precipitation), which can contribute to the change in character above and below the jovicline.  Maybe both of these processes are at work and intricately intertwined.

Irrespective of the interpretation, Juno's microwave radiometer has revealed that a significant transition in the microwave brightness of Jupiter's mid-latitude belts and zones (associated with ammonia, temperature, or both) occurs in the 5-10 bar region, and we hope that future studies will allow us to explain its origins.

\acknowledgments
Fletcher is a Juno Participating Scientist supported by a Royal Society Research Fellowship and European Research Council Consolidator Grant (under the European Union's Horizon 2020 research and innovation programme, grant agreement No 723890) at the University of Leicester.  Orton is supported by funds from NASA distributed to the Jet Propulsion Laboratory, California Institute of Technology.   Some of this research was carried out at the Jet Propulsion Laboratory, California Institute of Technology, under a contract with the National Aeronautics and Space Administration (80NM0018D0004).  Wong is supported by NASA's Juno Participating Scientist program through grant 80NSSC19K1265 to SETI Institute.  Kaspi, Galanti and Duer are supported by the Minerva Foundation and the Helen Kimmel Center for Planetary Science at the Weizmann Institute of Science.  Guillot is supported by a grant from the Centre National d’Etudes Spatiales.  We are grateful to J. Rogers for helpful insights into features in Jupiter's STB and NTB, T. Dowling and J. Aurnou for insights on deep temperature gradients, and to two anonymous reviewers for helping to improve the quality of this article. Juno observations are available through the Planetary Data System Atmospheres Node (\url{https://pds-atmospheres.nmsu.edu/data_and_services/atmospheres_data/JUNO/microwave.html}), and links to the specific calibrated MWR data (\url{https://pds-atmospheres.nmsu.edu/PDS/data/jnomwr_1100/}).  Data for individual figures are available through Zenodo (\url{https://doi.org/10.5281/zenodo.4761404}).

\appendix
\label{app}

\section{Cassini Winds vs. Hubble Winds}
\label{appS1}

\begin{figure*}
\begin{center}
\includegraphics[angle=0,width=1.1\textwidth]{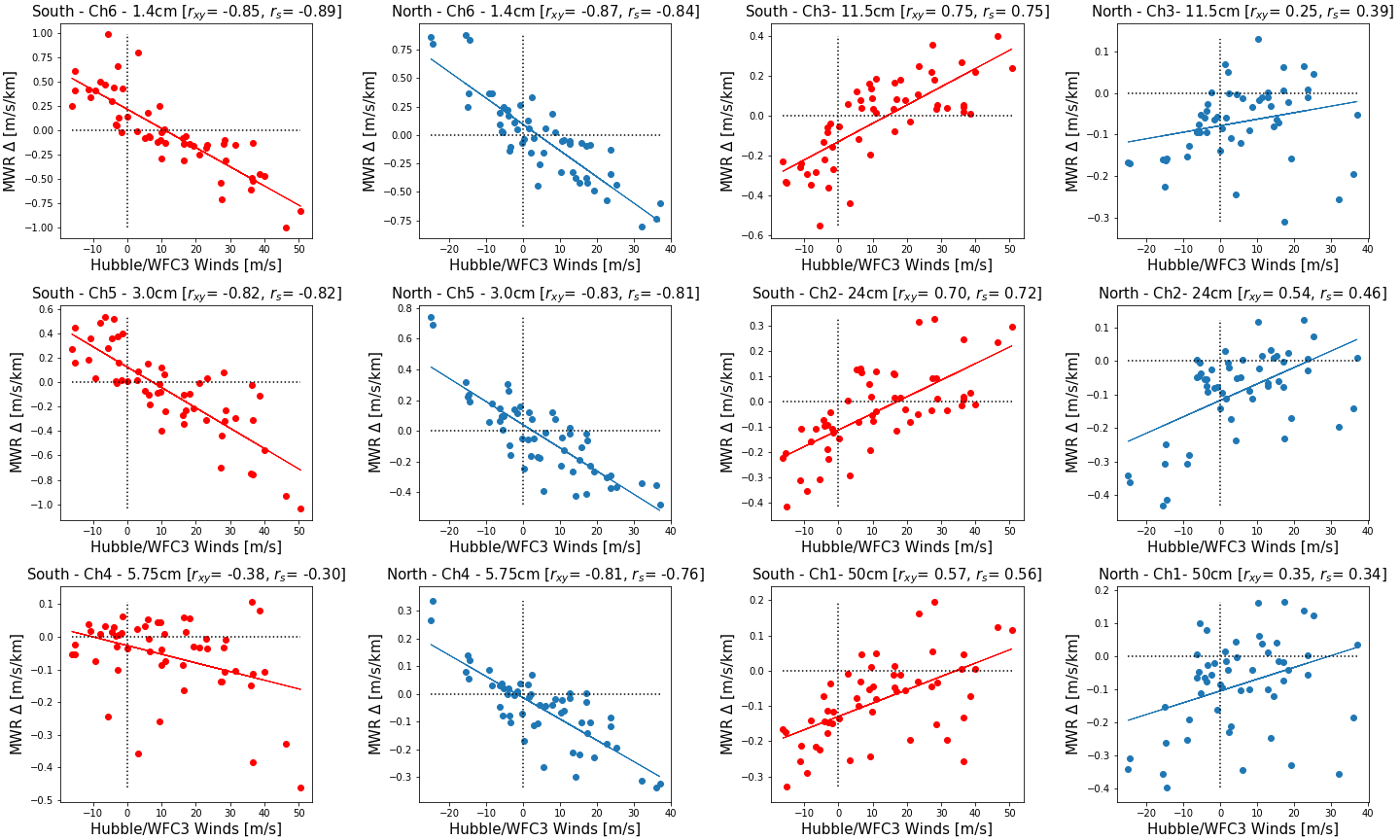}
\caption{Scatter plots revealing negative (channels 4-6, left columns) and positive (channels 1-3, right columns) correlations between the nadir microwave $T_B$ gradients $\Delta$ and the Hubble cloud-tracked winds of \citeA{17tollefson} for February 2017.  Only latitudes between $25^\circ$ and $65^\circ$ in each hemisphere are included.  Southern-hemisphere correlations are in red, northern-hemisphere correlations are in blue.  A linear trend line has been added as a guide. The Pearson $r_{xy}$ and Spearman's ranked $r_s$ correlation coefficients are provided for each channel and hemisphere.  }
\label{scatter2017}
\end{center}
\end{figure*}

\begin{figure*}
\begin{center}
\includegraphics[angle=0,width=1.1\textwidth]{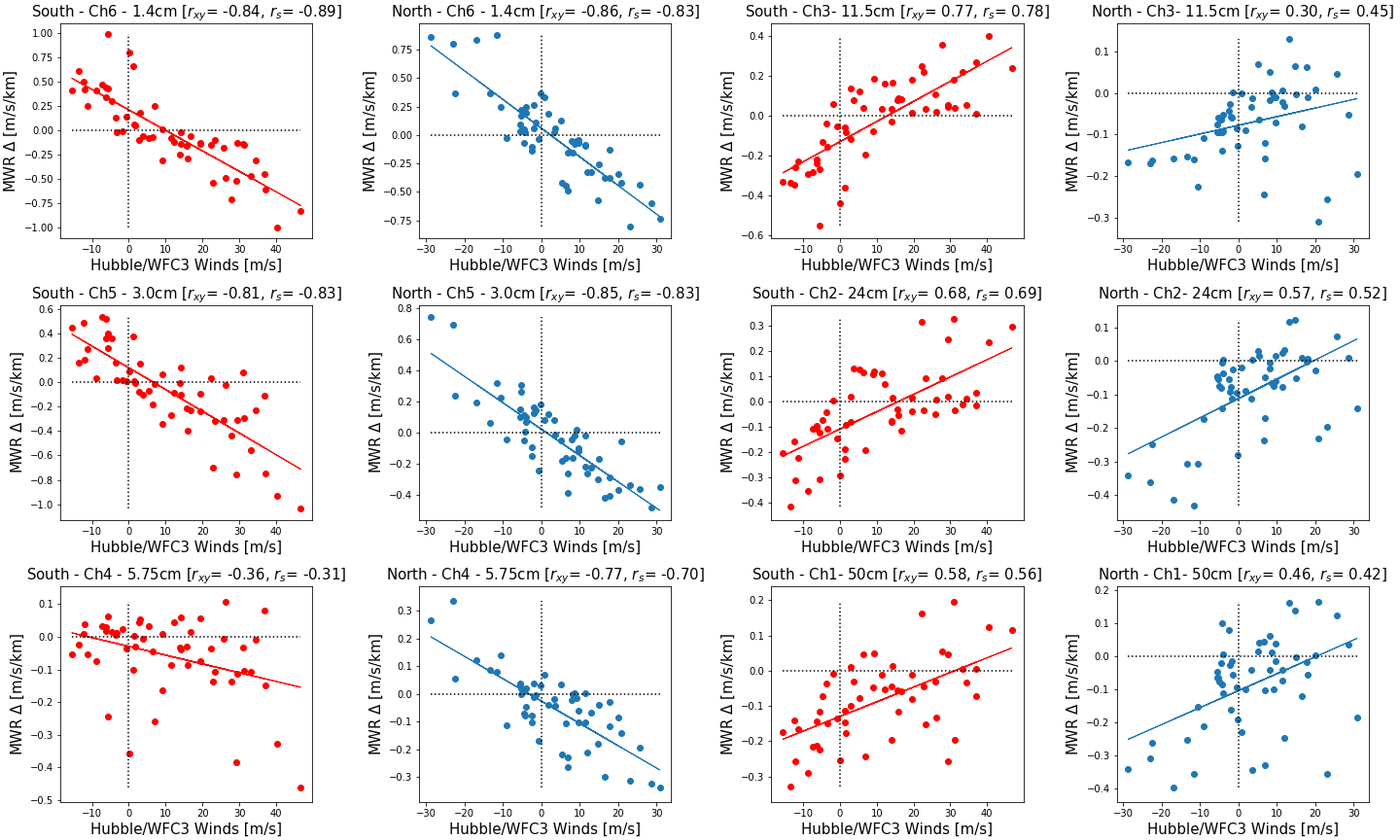}
\caption{Scatter plots revealing negative (channels 4-6, left columns) and positive (channels 1-3, right columns) correlations between the nadir microwave $T_B$ gradients $\Delta$ and the Hubble cloud-tracked winds of \citeA{20wong} for June 2019.  Only latitudes between $25^\circ$ and $65^\circ$ in each hemisphere are included.  Southern-hemisphere correlations are in red, northern-hemisphere correlations are in blue.  A linear trend line has been added as a guide. The Pearson $r_{xy}$ and Spearman's ranked $r_s$ correlation coefficients are provided for each channel and hemisphere.  }
\label{scatter2019}
\end{center}
\end{figure*}

The main article uses cloud-top zonal winds derived from Cassini/ISS measurements \cite{03porco} during the 2000 Jupiter flyby.  In particular, Fig. \ref{scatter} showcases the strength of the Pearson and Spearman correlation between the microwave brightness gradients ($\Delta$) and the cloud-top winds, showing how the results fall naturally into two groups:  positive correlation in deep-sounding channels 1-3 (11.5-50 cm), and negative correlation in shallow-sounding channels 4-6 (1.4-5.75 cm).  In Figs. \ref{scatter2017} and \ref{scatter2019} we instead use cloud-top winds derived in 2017 \cite{17tollefson} and 2019 \cite{20wong} from Hubble Space Telescope WFC3 observations.  Although the zonal winds have remained rather stable over time, small changes in the windspeeds do occur, and can have a small effect on the strength of the correlations.  The Pearson correlation coefficients for each hemisphere, wavelength, and zonal wind profile are compared in Table \ref{S1}, where we find small improvements in the strength of the positive and negative correlations when the Hubble winds are used, but no alteration to the conclusions of our study in the main article.  For completeness, Table \ref{S2} also provides the $p$-values for each correlation, with values significantly smaller than 0.05 allowing us to firmly reject the null hypothesis that the winds and the MWR gradients are uncorrelated.  The $p$-values only come close to this limit in Channels 4 (southern hemisphere) and 3 (northern hemisphere) where the correlation coefficient tends to zero near the jovicline.

\begin{table}
\caption{Comparing Pearson correlation coefficients between nadir microwave brightness gradients $\Delta$ and cloud-top zonal winds from Cassini in 2000 \cite{03porco} and Hubble in 2017 \cite{17tollefson} and 2019 \cite{20wong}.}
\centering
\begin{tabular}{l l l l l}
\hline
 Channel  & Wavelength (cm) & Cassini-2000 $r_{xy}$ & Hubble-2017 $r_{xy}$ & Hubble-2019 $r_{xy}$  \\
\hline
South & & & & \\
\hline
1 & 50 & 0.545 & 0.566 & 0.581\\
2 & 24 & 0.673 &  0.697 & 0.680 \\
3 & 11.5 & 0.754 & 0.754 &  0.775 \\
4 & 5.75 & -0.271 & -0.382 & -0.361 \\
5 & 3.0 & -0.741 & -0.816 & -0.813 \\
6 & 1.4 & -0.820 & -0.847 & -0.845 \\
\hline
North & & & &  \\
\hline
1 & 50 & 0.455 & 0.351 & 0.459 \\
2 & 24 & 0.559 & 0.539 & 0.575 \\
3 & 11.5 & 0.340 & 0.249 & 0.300 \\
4 & 5.75 & -0.720 & -0.808 & -0.771 \\
5 & 3.0 & -0.814 & -0.831 & -0.850 \\
6 & 1.4 & -0.821 & -0.866 & -0.861 \\
\hline
\end{tabular}
\label{S1}
\end{table}

\begin{table}
\caption{Comparing Pearson correlation p-values between nadir microwave brightness gradients $\Delta$ and cloud-top zonal winds from Cassini in 2000 \cite{03porco} and Hubble in 2017 \cite{17tollefson} and 2019 \cite{20wong}.}
\centering
\begin{tabular}{l l l l l}
\hline
 Channel  & Wavelength (cm) & Cassini-2000 $p_{xy}$ & Hubble-2017 $p_{xy}$ & Hubble-2019 $p_{xy}$  \\
\hline
South & & & & \\
\hline
1 & 50 & $2.06\times10^{-05}$ & $8.21\times10^{-06}$ & $4.11\times10^{-06}$\\
2 & 24 & $2.46\times10^{-08}$ & $4.67\times10^{-09}$ & $1.54\times10^{-08}$ \\
3 & 11.5 & $4.70\times10^{-11}$ & $4.52\times10^{-11}$ & $6.14\times10^{-12}$ \\
4 & 5.75 & $4.72\times10^{-02}$ & $4.35\times10^{-03}$ & $7.40\times10^{-03}$ \\
5 & 3.0 & $1.47\times10^{-10}$ & $5.97\times10^{-14}$ & $8.54\times10^{-14}$ \\
6 & 1.4 & $3.33\times10^{-14}$ & $6.7\times10^{-16}$ & $9.85\times10^{-16}$ \\
\hline
North & & & &  \\
\hline
1 & 50 & $5.52\times10^{-04}$ & $9.33\times10^{-03}$ & $4.87\times10^{-04}$ \\
2 & 24 & $1.12\times10^{-05}$ & $2.58\times10^{-05}$ & $5.49\times10^{-06}$ \\
3 & 11.5 & $1.19\times10^{-02}$ & $6.96\times10^{-02}$ & $2.76\times10^{-02}$ \\
4 & 5.75 & $8.32\times10^{-10}$ & $1.53\times10^{-13}$ & $9.46\times10^{-12}$ \\
5 & 3.0 & $7.07\times10^{-14}$ & $7.79\times10^{-15}$ & $4.10\times10^{-16}$ \\
6 & 1.4 & $2.79\times10^{-14}$ & $2.76\times10^{-17}$ & $7.20\times10^{-17}$ \\
\hline
\end{tabular}
\label{S2}
\end{table}

\section{Robustness of Correlations}
\label{appS2}

\begin{figure*}
\begin{center}
\includegraphics[angle=0,width=0.75\textwidth]{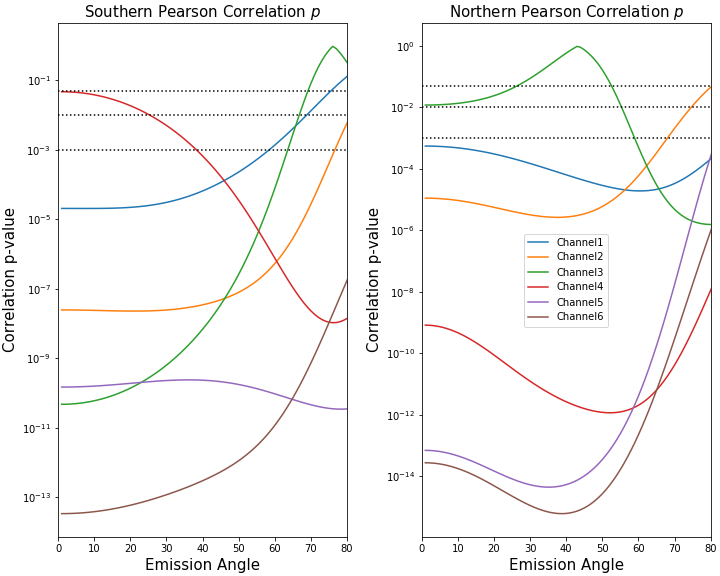}
\caption{Variation of the $p_{xy}$ values with emission angle for the Pearson correlation coefficients $r_{xy}$ between Cassini winds and MWR brightness gradients shown in Fig. \ref{pearson} of the main text.  Horizontal lines at 0.05, 0.01 and 0.001 show the thresholds for statistical significance (values below 0.05 allowing rejection of the null hypothesis that the winds and MWR brightnesses are uncorrelated). }
\label{pearsonp}
\end{center}
\end{figure*}

The main text uses a weighted average over nine perijoves of the coefficients used to define the limb darkening observed by MWR in each channel, using the techniques of \citeA{20oyafuso}.  Fig. \ref{pearson} shows the variation with emission angle of the Pearson correlation coefficient $r_{xy}$ between cloud-tracked winds from Cassini and microwave brightness gradients.  Fig. \ref{pearsonp} shows the how the corresponding $p_{xy}$ values vary with emission angle, confirming that the null hypothesis (that winds and microwave gradients are uncorrelated) can be rejected everywhere except in MWR channel 3 for emission angles of $\sim45^\circ$ in the north and $\sim75^\circ$ in the south.  These are the transition points, where the correlation flips from positive to negative, and are assigned to the 5-14 bar range in the main text.

\begin{figure*}
\begin{center}
\includegraphics[angle=0,width=0.75\textwidth]{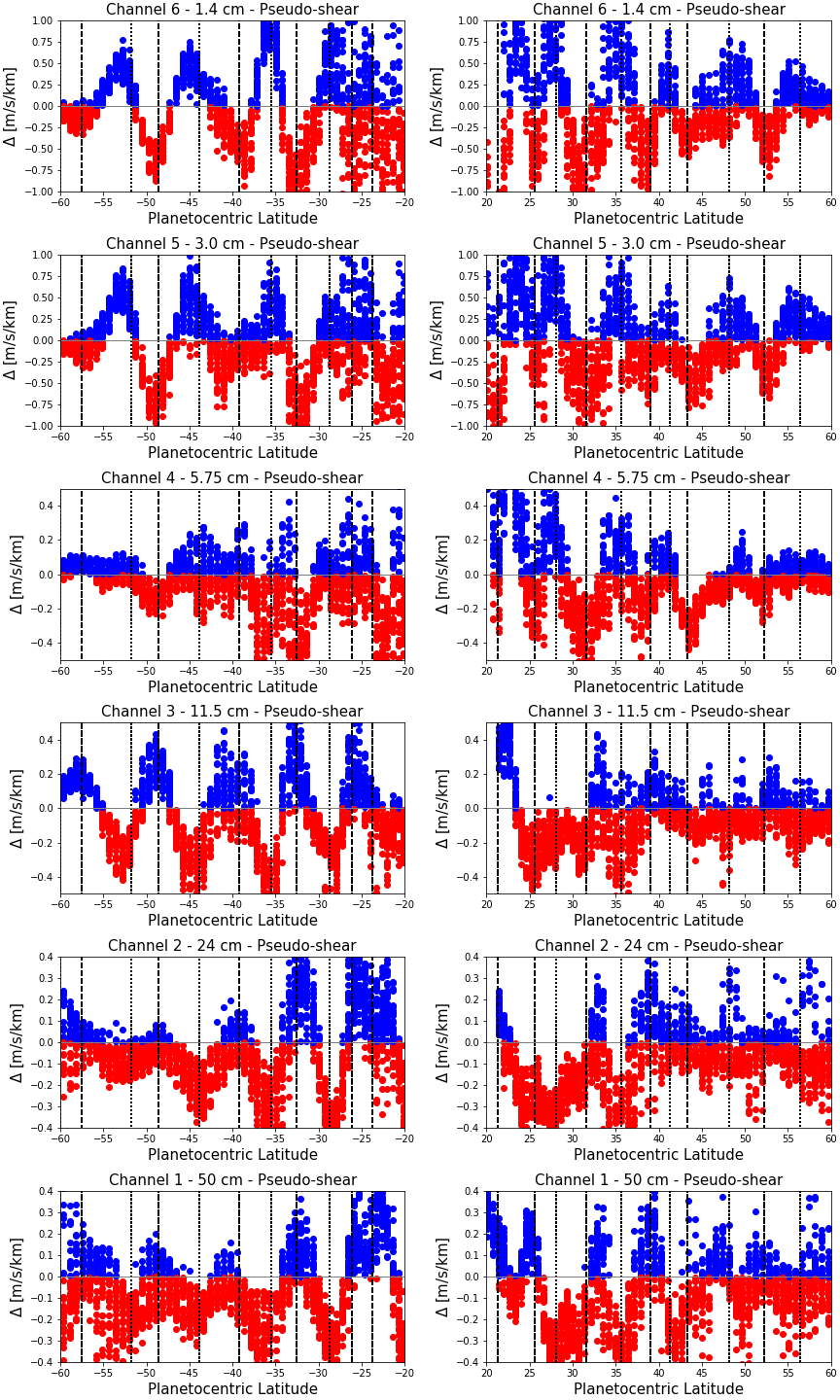}
\caption{Assessing the robustness of the MWR brightness temperature gradients from the main paper, where a weighted average of the nadir $c_0$ coefficient over nine perijoves was used to compute $\Delta$.  Here we take 36 combinations of two perijoves from the nine, and recompute $\Delta$ for each pair.}
\label{dudz_robust}
\end{center}
\end{figure*}

We can also question whether the measured pseudoshears and correlations can change if we use different combinations of perijoves, rather than all nine.  In this section, we explore the robustness of the correlations.  From nine different perijoves there are 36 different combinations of two perijoves.  For each of these 36 pairs, we compute the mean nadir brightness temperature and the associated pseudo-windshear $\Delta$, shown as the scatter of points at each latitude in Fig. \ref{dudz_robust}.  The same basic structure of peaks in $\Delta$ coinciding with peaks in the cloud-tracked winds persists for all 36 combinations, although the scatter becomes larger for the deep-sounding channels 1 and 2 (24-50 cm).  This is consistent with the fact that \citeA{17ingersoll} were able to identify the switch in the brightness of the belts even from a single perijove (PJ1).  

\begin{figure*}
\begin{center}
\includegraphics[angle=0,width=1.1\textwidth]{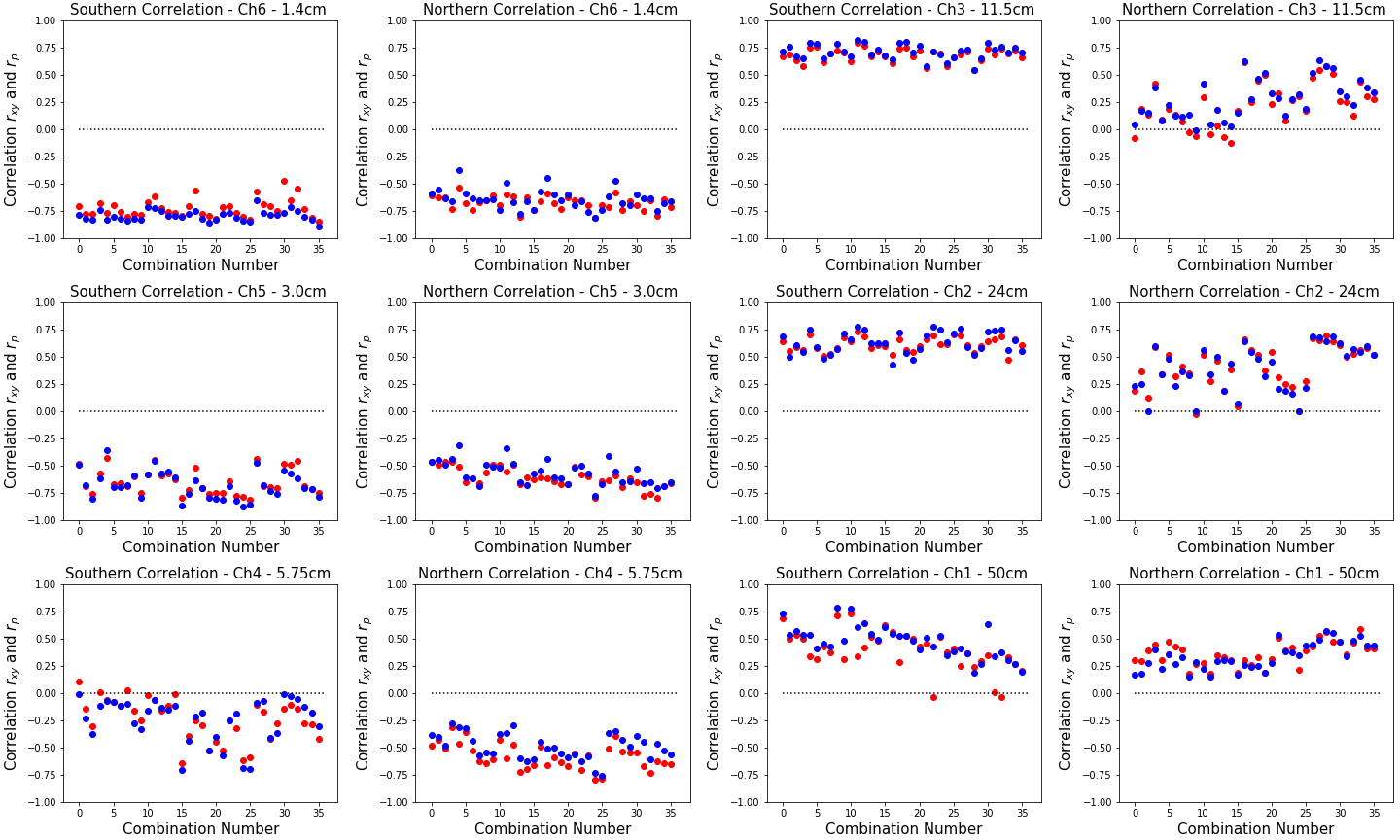}
\caption{The Pearson and Spearman correlation coefficients (red and blue, respectively) for each of the 36 pairs of perijoves, confirming that the correlations identified in the main article from a weighted average of nine perijoves are robust.}
\label{correlation_robust}
\end{center}
\end{figure*}

\begin{figure*}
\begin{center}
\includegraphics[angle=0,width=1.1\textwidth]{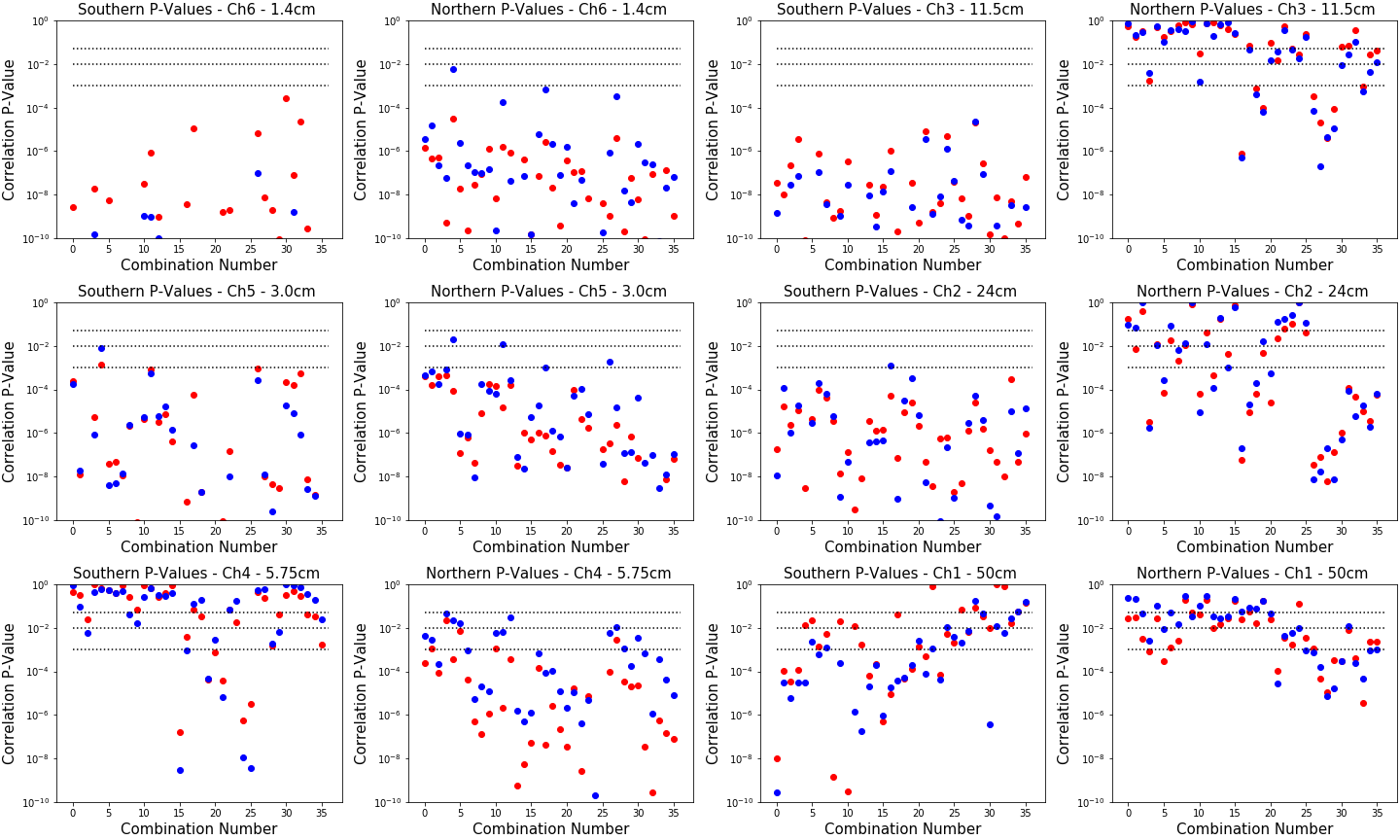}
\caption{Probability values for the Pearson and Spearman correlation coefficients (red and blue, respectively) for each of the 36 pairs of perijoves in Fig. \ref{correlation_robust}.  Any dot falling below the topmost horizontal line at 0.05 is considered to be statistically significant, dots falling below the lower horizontal line at 0.001 is considered highly statistically significant. }
\label{prob_robust}
\end{center}
\end{figure*}

Next, we recompute the Pearson and Spearman correlation coefficients for each of the 36 combinations in Fig. \ref{correlation_robust}, along with the associated probability values in Fig. \ref{prob_robust}.   Each pair of perijoves confirms that the correlation between $\Delta$ and the cloud-tracked winds falls into two distinct categories (negative correlation for shallow-sounding channels, positive correlation in deep-sounding channels), and show that correlation is weak in the southern hemisphere for Channel 4, and weak in the northern hemisphere for Channel 3.  The strength of the correlation does change depending on which PJ pairs are considered, which is why the main article uses a weighted average over all nine.  The p-values reveal a similar story - if the p-value is considerably smaller than 0.05 (the topmost dotted horizontal line), then the correlation is statistically significant and we can firmly reject the null hypothesis of zero correlation between the MWR $\Delta$ and the winds.  This is the case for shallow-sounding channels 5 and 6, and for deep-sounding channels 2 and 3 in the south.  Fig. \ref{prob_robust} also confirms that zero correlation is found in channel 4 (south) and channel 3 (north), where the transition is occurring.  However, we find that channel 1 p-values in some PJ pairs are sufficiently large that we cannot reject the null hypothesis (some exceeding $p=0.05$).  This implies that, if we only had observations from two perijoves, the correlation in channel 1 would be hard to see, again consistent with \citeA{17ingersoll}, who could see the correlation in the 40-60 bar range with PJ1 data, but not deeper.  We conclude that the detection of the correlation with statistical significance at the deepest levels sensed by MWR is only made possible via the weighted average of more than two perijoves.

\section{Alternative Contribution Functions}
\label{appS3}

\begin{figure*}
\begin{center}
\includegraphics[angle=0,width=0.8\textwidth]{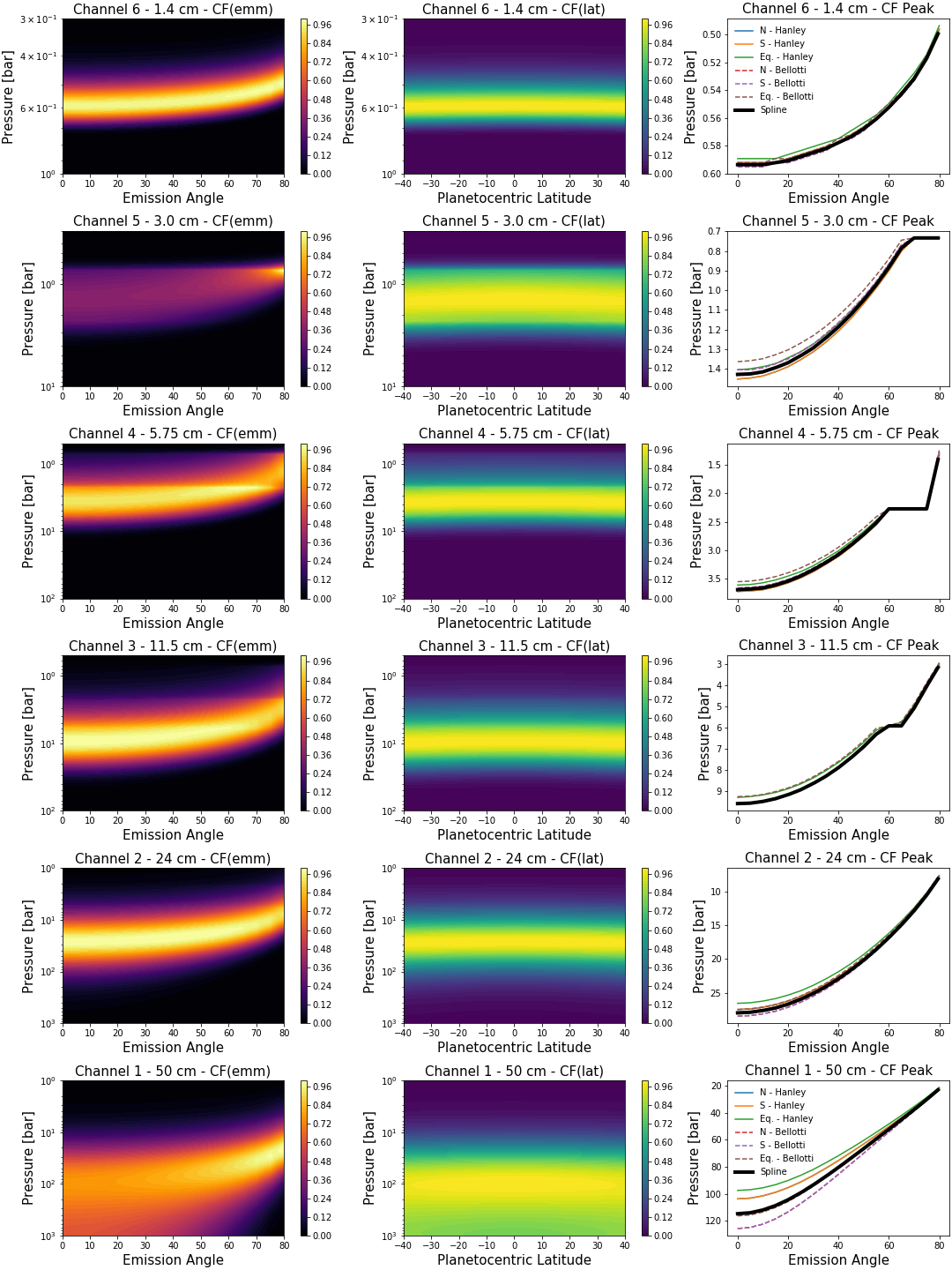}
\caption{Contribution functions based on a deep $2.76\times$ enrichment of NH$_3$, decreasing with height as the different cloud layers condense.  Left:  normalised contribution functions as a function of emission angle for the equator.  Centre: normalised contribution functions at zero emission angle (nadir view) for all latitudes.  Right: peak pressure of the contribution function averaged over three regions (north $20^\circ$N to $40^\circ$N; south $20^\circ$S to $40^\circ$S; and equator $5^\circ$N to $5^\circ$S) using two different NH$_3$ opacity models - \citeA{09hanley} as the solid lines and \citeA{16bellotti} as the dashed lines. The solid black line is the spline-interpolated contribution function described in the main text.  }
\label{contfn_default}
\end{center}
\end{figure*}

In the main text, we point out that the depths sounded by each MWR channel can only be estimated using a model - specifically, the vertical distribution of gases like ammonia and water, and models for their spectral opacity.  In the main article we use the current best estimates of the latitudinal distribution of NH$_3$ \cite{20guillot_ammonia} to calculate contribution functions to assign pressure ranges to each MWR channel.  As a demonstration of model dependence, Fig. \ref{contfn_default} recomputes these contribution functions using a more simplistic NH$_3$ distribution.  Namely, we assume NH$_3$ to be uniform in latitude, to have a deep abundance equivalent to $2.76\times$ the solar NH$_3$ abundance \cite{17li}, and to decrease with altitude as it condenses into cloud decks assuming thermochemical equilibrium.  These assumptions are typically used to report the altitude sensitivity of MWR \cite{17janssen}, and a comparison of Fig. \ref{contfn_default} with Fig. \ref{contfn} of our main article shows how the peak pressure levels change.  Specifically, the depth sounded at nadir changes from $\sim5$ bar to $\sim3.6$ bar in Channel 4, and from $\sim14$ bar to $\sim9.5$ bar in Channel 3, if we use the crude NH$_3$ vertical distribution, which would have the effect of moving the transition level to slightly shallower pressures.   Sharp kinks in the contribution occur where NH$_3$, NH$_4$SH and H$_2$O condense with these assumptions.  We expect the contribution functions reported in the main article to be a more accurate representation of the altitude sensitivity, given that they incorporate the measured NH$_3$ depletion down to the 40-60 bar level \cite{17li}.

\section{Calculation of Gravity and Height}
\label{appS4}

\begin{figure*}
\begin{center}
\includegraphics[angle=0,width=1.0\textwidth]{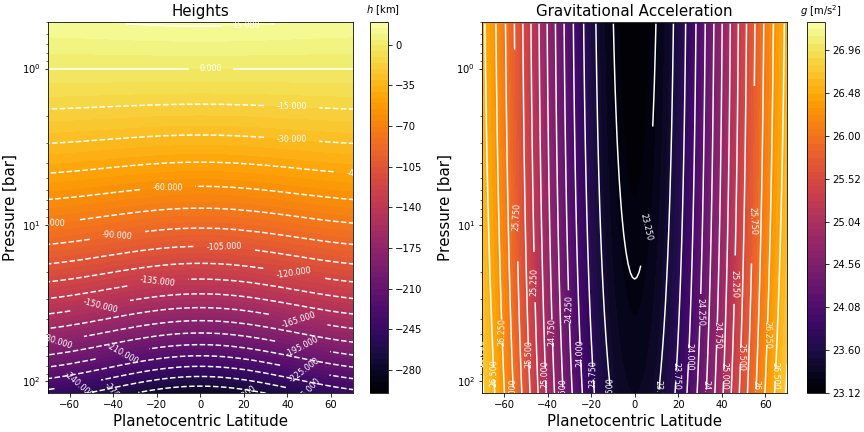}
\caption{Estimates of Jupiter's height and gravitational acceleration for use in the estimates of pseudo-shears and integrated zonal winds.  We used the combined gravitational and centrifugal potential of \citeA{20buccino} to estimate the effective gravity $g(\phi,p)$, reproducing their 1-bar gravitational acceleration.  The ideal gas law then allows us to calculate the depths $z(\phi,p)$, which matches those measured by the Galileo probe \cite{98seiff}.}
\label{gravity_calc}
\end{center}
\end{figure*}

In order to integrate the thermal wind equation as a function of altitude, and to adjust for gravitational acceleration in our measurement of microwave brightness gradients, we must match our pressure grid to an appropriate altitude grid $z(\phi,p)$ and gravitational acceleration $g(\phi,p)$.  Fig. \ref{gravity_calc} shows our two grids, calculated using the gravitational and centrifugal potential of \citeA{20buccino} and the ideal gas law, cross-checking the estimated heights with those measured by the Galileo probe \cite{98seiff}.

%
%

\bibliography{references.bib}

%
%
%
%
%

\end{document}